\documentclass[a4paper,12pt]{article}
\usepackage{jcappub} 
\usepackage{cite}
\usepackage{graphicx}
\usepackage{enumerate}
\usepackage{hyperref}

\usepackage{cancel}
\usepackage{color}
\usepackage{graphicx}
\usepackage{amsmath}
\usepackage{amsfonts}
\usepackage{amssymb}
\usepackage{rotating}
\usepackage{booktabs}
\usepackage{xcolor}
\usepackage{soul}

\def\blue#1{\textcolor{blue}{#1}}
\def\green#1{\textcolor{green}{#1}}
\def\orange#1{\textcolor{orange}{#1}}

\def\comment#1{}

\usepackage{graphicx}

\title{\boldmath 
Collimated and spinning fireballs for ultra-relativistic jets:\\
long vs short Gamma-ray bursts by angular momentum and mass ratio}

\author{She-Sheng Xue}

\affiliation{ICRANet Piazzale della Repubblica, 10 -65122, Pescara, Italy
\\ Physics Department, Sapienza University of Rome, 
Rome, Italy\\INFN, Sezione di Perugia, 
Perugia, Italy
\\ICTP-AP, University of Chinese Academy of Sciences, Beijing, China
}

\emailAdd{xue@icra.it, she-sheng.xue@cern.ch} 

\abstract{In this study, we investigate the gravitational collapses of rotating stellar systems accounting for Gamma-Ray Burst jet progenitors. Based on the virial theorem of hadron collisional relaxations and Newtonian slow-rotating approximation, we analyze the conversion of gravitational binding energy into kinetic energy of hadrons, whose collisions produce photons and electron-positron pairs forming fireballs. Our qualitative analysis implies that rotation effects collimated and spinning fireballs with nontrivial angular momenta along the propagating direction, thus making 
ultra-relativistic jets. Results reveal the possible trends that the fireball becomes more collimated and the jet angle decreases as the total angular momentum and mass ratio $J/M$ of the slow-rotating collapsing core increases. Discussing the extrapolation of these trends to fast-rotating collapsing systems, we speculate that the ratio $J/M$ should be a key quantity for differentiating long bursts (massive core collapses) from short bursts (binary coalescence). We derive the intrinsic correlations of collimated fireball quantities that should be imprinted on a large sample of observed GRB data as empirical correlations.
}

\begin{document}

\maketitle
\flushbottom

\newpage

\section{Introduction}\label{sec:0}

Gamma-ray bursts (GRBs) are the most energetic and complex events in the Universe. Within mere seconds, their progenitors release high-energy photons of tremendous 
isotropic equivalent energy estimated to range from $10^{49}$ to $10^{54}$ ergs, assuming progenitors originate from opaque spherical fireballs of large particle number and energy densities \cite{Rees:1992ek}. The total energy budget should be smaller, considering that GRBs are highly collimated events, with their energy confined within a narrow ultra-relativistic collimated outflow (jet) \cite{Meszaros_1998, Lamb_2021}. The typical jet opening angle is usually a few degrees \cite{Wang_2018, Mizuta_2013}. For example, the jet opening angle of GRB 170817A is estimated to be around 20 degrees \cite{He_2018, Lamb_2018}. The jet opening angle of the brightest GRB 221009A is estimated to be exceptionally narrow, approximately 0.6 degrees \cite{2023arXiv230301203A,2024JHEAp..41...42Z}.  

Significant advancements in our comprehension of GRBs dynamics and phenomena have been achieved through extensive research efforts \cite{2004RvMP...76.1143P,2006RPPh...69.2259M,2014ARA&A..52...43B,2015JHEAp...7...73D,2015PhR...561....1K,zhang_2018, Ruffini2019
}. 
Among others, the ultra-relativistic hydrodynamical outflow 
of spherical fireballs has been studied well numerically and analytically, e.g., Refs.~\cite{2004RvMP...76.1143P,zhang_2018, 2012grb..book.....K, Ruffini1999, Ruffini2000, Meszaros2000, Zhang2021}. 
Recently, numerical investigations have made notable progress in understanding jet evolution from early launching to later propagation and interaction with ambient stellar medium away from jet 
progenitors, e.g.,~Refs.~\cite{Kiuchi2024, Xie2024,Kiuchi2023, Fujibayashi2023,Gottlieb2022, Gottlieb2021,2021MNRAS.502.1843N,2018ApJ...863...58X,2021MNRAS.500..627H,2011MNRAS.418L..79T}.

However, what are the dominant dynamics of GRB progenitors that ultimately yield the formation of an energetic and optically thick plasma composed of photons and electron-positron pairs, dubbed a collimated and spinning fireball, making an ultra-relativistic jet? 
There are numerous theoretical investigations in the direction following the Penrose process \cite{1969NCimR...1..252P, Penrose1971} of extracting rotational energy from particles falling into a spinning black hole (SBH), and Blandford and Znajek's subsequent work \cite{1977MNRAS.179..433B,1982MNRAS.199..883B} that jet energy could be extracted from the SBH rotational energy and transported out along large-scale magnetic fields surrounding SBH, making a relativistic jet. 

In contrast and comparison, we study the possibility of how significant amounts of stellar gravitational binding 
energy and angular momentum are released to collimated and spinning fireballs in the gravitational collapse and coalescence of massive and rotating stellar systems.
The critical questions are within just a few seconds during gravitational collapses: (i) how this immense gravitational energy is converted (extracted) into photon and pair energies; (ii) why photon and pair energies form opaque and collimated fireballs of huge energy and number densities; (iii) how collimated fireballs acquire nontrivial angular momentum and make ultra-relativistic jets. 
In Ref.~\cite{Xue_2021}, we study how the spherical fireball forms during gravitational homologous collapses of massive stellar cores. 
Here, considering the collapses of massive and rotating stellar systems, we study the formation of a collimated fireball with nontrivial angular 
momentum, whose hydrodynamical outflow makes an ultra-relativistic jet. 

In a self-gravitating system, Lynden-Bell and Wood \cite{LyndenBell1967, LyndenBell1968} pioneered the study of gravothermal catastrophe using the framework of violent hadron collisionless relaxation and the virial theorem. Their work demonstrated the relaxation process in which hadrons gain their mean kinetic heat energy from gravitational potential energy. 
Based on the local virial theorem and the homologous core collapse scenario \cite{Goldreich1980}, we show that hadrons gain mean kinetic heat energy from gravitational binding energy through collisional relaxation. On the other hand, the production of photons by hadron collisions has been extensively studied in the context of heavy-ion collisions \cite{Turbide2004, Kapusta1991, Kapusta1992, Arnold2001, Gale2015, Hidaka2015}. These two aspects imply that hadron-hadron collisions subsequently produce photons and pairs, converting their mean kinetic energy to photons' energy. As a result, the gravitational binding energy converts into the energies of opaque photon-pair plasma (spherical fireball) in gravitational collapses \cite{Xue_2021}. 
This article investigates the possibility of forming collimated fireballs and ultra-relativistic jets in gravitational homologous collapses of massive rotating core collapses and binary coalescence. 

We organize this article as follows. Based on the adiabatic assumption, Sections \ref{adi}, \ref{virial}, and \ref{hadron} present discussions on gravothermal catastrophe, the virial theorem, and photon production from hadron relaxations and collisions. In Section \ref{Homo}, we discuss the gravitational homologous collapse of a slowly rotating core. Sections \ref{psphere} and \ref{jetf} study collimated fireball and 
ultra-relativistic jet properties. 
Section \ref{corre} discusses the intrinsic correlations of collimated fireball properties and how they relate to the empirical correlations from GRB observations. 
Section \ref{spherevsjet} discusses the trends of collimated fireball properties as varying progenitor angular momentum and mass ratio, in connection with the characteristics of long versus short GRBs.

The stellar core density is characterised by the nuclear saturation density $n_0\approx 0.16 \,{\rm fm}^{-3}=1.6\times 10^{38}/{\rm cm}^3$ or $\rho_0=mn_0\approx 2.4\times 10^{35} {\rm ergs}/{\rm cm}^3$, and $m\approx 1$ GeV is the typical baryon (neutron) mass. The light speed $c=1$, Planck constant $\hbar =1$, and Boltzmann constant $k=1$ are used unless otherwise specified.  

\section{Adiabatic approximation in gravitational collapse}\label{adi}

Gravitational collapse is a dynamic avalanche involving complex microscopic and macroscopic processes at vastly different scales. It is 
challenging for both analytical and numerical approaches to analyze all these dynamics without any approximations. 
To make appropriate approximations, we need to understand the dominant physical processes 
and their variations across the relevant length and time scales.

We consider a rotating stellar core of mass $M$, radius $R$, 
and angular velocity $\Omega$ undergoing a gravitational collapse process.  
The most relevant timescales $\tau_{\rm grav}$ are collapsing $|R|/|\dot R|$, rotating $1/\Omega$ and hydrodynamics timescale $\sim R/v_s$, where 
$v_s$ is the sound velocity. These macroscopic scales are much larger than the microscopic scales of baryon 
collisional relaxation, photon production, and thermalization.

This disparity in scales implies that:
\begin{itemize}
    \item  these microscopic processes 
occur not only ``instantaneously'' but also ``locally'' 
compared with gravitational collapsing 
and hydrodynamic processes.
\item There is no causal correlation among the concurrent microscopic processes 
at different space-time points on macroscopic scales.
\end{itemize}

In other words, the vast difference between microscopic and macroscopic timescales suggests that gravitational collapse can be considered a very slow, {\it adiabatic} process compared to these rapid and local microscopic processes. Therefore, these microscopic processes can be approximately analyzed as if the self-gravitating cores were static.

Based on this adiabatic approximation, we discuss local thermal equilibrium configurations during gravitational collapses.

\section{Local virial theorem and gravothermal configuration }\label{virial}

We consider a stellar core composed of $N$ baryons of mass $m$\footnote{By the term {\it baryons}, we indicate nucleons, nuclei, hadrons, quark-gluon plasma that carry baryon numbers.}. 
In gravitational collapse, baryons gain their
mean kinetic energy $F$ from the gravitational binding energy $U$ via their collisionless relaxation in time-varying gravitational potential. 
This gravothermal catastrophe phenomenon of stellar cores is studied by considering violent relaxation \cite{LyndenBell1967} and 
equipartition theorem \cite{LyndenBell1968}. 
In these studies, authors considered the stellar core to be an isothermal core of baryon temperature $T\ll m$ and internal ``heat" energy $F=\frac{3}{2}TN\ll mN$. 
It is in an equilibrium or equipartition configuration, 
where the Clausius virial theorem applies,
\begin{eqnarray}
2F+ U + U_c=3 PV,
\label{gravP} 
\end{eqnarray}
where $V$ is the core volume, and $P$ is the external pressure acting on the core surface. Here, $U_c \ll U$ represents the centrifugal potential energy for a slowly rotating core, which we will further justify.

These studies indicate that the baryon collisionless relaxation process in 
the gravitational potential and under external pressure leads to two key consequences: 
\begin{itemize}
    \item   The gravitational energy is converted to the internal heat energy of baryons.
\item The balance between gravitational dynamics and thermodynamics is established, 
making the virial theorem (\ref{gravP}) applicable.
\end{itemize} 
The second point holds, provided that the time scales of microscopic relaxation processes are much smaller than those of macroscopic gravitational and hydrodynamical processes, justifying the adiabatic approximation.
Following the discussions in the pioneer works \cite{LyndenBell1975a, Shapiro1977a}, 
it is noted that the gravitational evolution time scale due to stellar evaporation and consumption is longer than the core relaxation time scale, the core will maintain an approximate isothermal core profile (\ref{gravP}) and evolve homologously. 

We generalise \cite{Xue_2021} these discussions to a small fluid element of volume $dV$ inside the stellar core by considering a thermal equilibrium locally achieved through the relaxation processes of baryon-baryon collisions at nuclear density. We estimate the baryon collisional relaxation timescale as
\begin{eqnarray}
\tau_{\rm relax}=(\sigma_n v n)^{-1}\approx 5.21\times 10^{-3} (n_0/n) (T/m)^{1/2}T^{-1},
\label{relax} 
\end{eqnarray}
where $n=\rho/m$ ($v$) is the baryon number 
density (mean velocity), $\sigma_n\approx 1/m_\pi^2$ is the typical cross-section of baryon-baryon collisions, 
and averaged collision energy (baryon temperature) $T=(1/2)mv^2$. For $T\ll m$,
the baryon collisional relaxation time scale $\tau_{\rm relax}$ (\ref{relax}) 
is much smaller than the time scales ($T^{-1}$) of baryon kinetic motion and collisionless relaxation \footnote{In 
Eq.~(\ref{relax}), for $T\sim 10^2$ MeV and $n\sim n_0$, 
the baryon kinetic motion time scale $T^{-1}\sim 10^{-23}$ seconds, and the baryon collisional relaxation time scale (\ref{relax}) $\tau_{\rm relax}\sim 10^{-26}$ seconds, i.e., 
$\tau_{\rm relax}\ll T^{-1}$.}. 
Therefore, a {\it local} thermal equilibrium
can establish, and
the Clausius virial theorem applies. 
Analogously with {\it global} virial theorem (\ref{gravP}), the {\it local}  virial theorem is given by
\begin{eqnarray}
2dF+ dU +dU_c =3 pdV,\label{dgravP}
\end{eqnarray}
and 
$dF=(3/2)T (\rho/m)dV$, $dU=(1/2)\rho \phi dV$ and $dU_c=\phi_c \rho dV$. Here baryon mass density $\rho=m n$ and number density $n$.
The terms $\phi$ and $\phi_c$ are gravitational and centrifugal potential 
in a rotating frame, respectively. 
For each fluid element of mass energy $\rho dV=mndV$, the baryon temperature $T$ characterizes 
the mean kinetic energy $dF$ of baryon motions and collisions, $dF\ll \rho dV$ for $T\ll m$. The fluid element's rotating energy is much smaller than its gravitational energy, i.e., $dU_c\ll |dU|$. All quantities in Eq.~(\ref{dgravP}) depend on the spatial point.
For an isothermal core, integrating {\it local} virial theorem (\ref{dgravP}) over the core volume $\int dV$, one obtains the {\it global} virial theorem (\ref{gravP}), using $\int pdV =PV$. This shows that the local equilibrium properties extend to the global equilibrium properties of the entire core under the adiabatic approximation.

Moreover, we approximately describe the internal pressure $p$ by the polytropic equation of state (EoS) 
\begin{eqnarray}
p = \kappa \rho ^\gamma,
\label{eos} 
\end{eqnarray}
with two parameters: mean thermal index $\gamma$ and 
coefficient $\kappa$.
As a result, combining the local virial theorem (\ref{dgravP}) and the EoS (\ref{eos}), we obtain the baryon temperature:
\begin{eqnarray}
3T/m &\approx & 3v_s^2 (1/\gamma) - (1/2) \phi -\phi_c,
\label{dgravT}
\end{eqnarray}
where $v_s$ is the sound velocity given by,
\begin{eqnarray}
v_s^2 &=&\partial p/\partial \rho =\gamma p/\rho.\label{soundv}
\end{eqnarray}
This relationship shows that through collisional relaxations, baryons gain heat energy (temperature) from the gravitational potential, which is effectively reduced by the centrifugal potential.

\section{Photon and $e^+e^-$ pair production from hadron collisions}\label{hadron}

The baryon-baryon collisions produce photons and pairs of light-charged leptons and quarks. 
Consequently, the mean kinetic heat energy of baryons is converted to the energy of photons, 
and pairs of other light-charged leptons and quarks.  
From the studies of heavy-ion collisions \cite{Turbide2004, Kapusta1991, Kapusta1992,
Arnold2001,Gale2015,Hidaka2015,Kapusta1991, Kapusta1992}, 
we learn about photon production in these collisions and can obtain the photon number and energy densities as follow \cite{Xue_2021},
\begin{eqnarray}
n_\gamma &\approx & \frac{4}{3}\frac{\alpha\alpha_s}{\pi}T^3\ln\left( 1+ \frac{2.9}{4\pi\alpha_s}\right),
\label{gammaN}\\
\rho_\gamma &\approx & \frac{8}{3}\frac{\alpha\alpha_s}{\pi}T^4\ln\left( 1+ \frac{2.9}{4\pi\alpha_s}\right),
\label{rhog}
\end{eqnarray}
where $\alpha=1/137$. The QCD coupling $\alpha_s\approx 0.5$ indicates that the photon production described by equation (\ref{rhog}) dominates other QED processes producing photons. 

The gravothermal catastrophe produces not only  heat energy for massive baryons, given by
$(3/2)T(\rho/m) dV$ (as in equation (\ref{dgravP})), but also thermal 
energy for relativistic particles, described by $\rho_\gamma dV$ (as in equation (\ref{rhog})).
For a baryon temperature $T$ on the order of ${\mathcal O}(10^2)$ MeV, as will be shown later, 
the photon number density $n_\gamma$ (\ref{gammaN}) and energy density $\rho_\gamma$ (\ref{rhog}) are very large. It implies that photons become opaque and thermalized by interacting with pairs of electrons and positrons. The local mean temperature $T_\gamma$ of thermalized photons can be estimated by equating 
photon thermal energy density $T^4_\gamma$ to the energy density $\rho_\gamma$ given in equation (\ref{rhog})
\begin{eqnarray}
\frac{\pi^2}{15}T_\gamma^4\approx \rho_\gamma,\quad
\Rightarrow \quad T_\gamma \approx 0.21 T,
\label{Tphoton}
\end{eqnarray}
because the local thermal equilibrium is approximately established. This result shows that the local mean photon temperature $T_\gamma$ is almost the same order as the local mean baryon temperature $T$. 

The fourth and back processes involving the production and annihilation of electron-positron pairs, 
$\gamma+\gamma\leftrightarrow e^++e^-$, are kinematically possible because the photon characteristic energy scale $T_\gamma$ (\ref{Tgamma}) is larger than electron and position masses $2m_e$. The cross-section $\sigma_{\gamma\gamma\leftrightarrow e^+e^-}$ 
is of the order of the Thomson cross-section 
\begin{eqnarray}
\sigma_\gamma \approx \sigma_T(3/8)(m_e/T_\gamma)\ln (2T_\gamma/m_e)\nonumber
\end{eqnarray}
and $\sigma_T=(8\pi/3)\alpha^2/m^2$.  
The corresponding photon mean-free length is given by
\begin{eqnarray}
\lambda_\gamma=(\sigma_\gamma n_\gamma)^{-1}\approx \frac{3.23\times 10^{6}}{[\ln (2T_\gamma/m_e)]} \left(\frac{m_eT_\gamma}{T^2}\right) \frac{1}{T} \sim {\mathcal O }(10^{-10})~ {\rm cm},  
\label{Tgamma}
\end{eqnarray} 
which is much smaller than a macroscopic size, namely optical depth $\tau_\gamma$ is very large.  

This indicates that via processes $\gamma+\gamma\leftrightarrow e^++e^-$, photons and electron-positron 
pairs participate in thermalisation or thermal equipartition in particle energy and number. 
As a result, these processes lead to an electrically neutral, deeply opaque, and thermalized plasma (fluid) of photons and electron-positron pairs, i.e., a ``collimated fireball". More detailed discussions can be found in Ref.~\cite{Xue_2021}.

As a consequence, the isothermal core profile is approximately described by 
baryon temperature $T$ (equation (\ref{dgravT})) and photon temperature $T_\gamma$ (equation (\ref{Tphoton})), as well as the photon production number and energy densities (equations (\ref{gammaN}) and (\ref{rhog})). The core will maintain these isothermal profiles and evolve homologously.

We add a few remarks to end this section. The 
photon number and energy densities (\ref{gammaN}) and (\ref{rhog}) and efficiency of producing such densities 
should be underestimated ({\it lower bounds}) in hadron collisions at densities $(\rho\sim 10\rho_0)$. From the timescales of high-energy photons' interacting dynamics and kinematics, we expect the exponential multiplicity of photon productions, like 
QED cascade \cite{Landau1938,
Elkina2011}, to occur via photon-photon and photon-electron collisions, 
as well as photons, charged particles and magnetic field (virtual photons) interactions. At higher densities $n\gtrsim 10n_0$, a phase transition from hadrons (baryons) to quark-gluon plasma should occur. In this case, the baryon temperature $T$ (\ref{dgravT}) should be the temperature of the quark-gluon plasma. We expect even more efficient photon production multiplicity via QCD cascade processes,  see Refs.~\cite{Fermi1950, Altarelli1977}, to occur in reality. In these high-energy photons' cascade processes, the total photon number $N_\gamma$ should be proportional to an exponential factor $e^{t^{\rm mac}_\gamma/t^{\rm mic}_\gamma}\gg 1$ and $t^{\rm mac}_\gamma> t^{\rm mic}_\gamma$, where $t^{\rm mic}_\gamma$ is the high-energy photon interacting 
timescale for particle productions 
and $t^{\rm mac}_\gamma$ is the timescale 
relating to the relaxation time (\ref{relax}) of hadron collisions in gravitational collapses. These topics deserve further investigation.

\section{Gravitational collapse of slowly rotating stellar cores}\label{Homo}

Observations of neutron stars, supernovas, and GRBs suggest that rotation is crucial in the core collapse scenario. However, realistic models for the final evolutionary state of a massive collapsing star that consistently includes rotation are still lacking. 
We proceed along the lines of previous Newtonian and rotating core collapse studies \cite{1985A&A...147..161E, 1997A&A...320..209Z}, which construct a set of differentially rotating polytropes in equilibrium as pre-collapse iron core models. 

\subsection{Euler and Poisson equations in Newtonian approximation}\label{newton}

We consider a slowly rotating stellar core undergoing a gravitational collapse process, whose rotation energy is much smaller than the gravitational binding energy. 
Following the semi-analytical approach \cite{Goldreich1980, Xue_2021}, we will study the homologous gravitational collapse of slowly rotating and massive spherical cores by using the gravitational and centrifugal potentials in Newtonian approximation.
We aim to explain how hadron collisions in gravitational potential, balanced by centrifugal potential, convert gravitational binding energy to the thermal energy of photons and pairs, creating a collimated fireball during gravitational collapses. Microscopic physical processes, e.g., photon production by hadron collisions,  at a small spacetime scale, are not affected much by using the general relativity (GR) or Newtonian framework at a macroscopic spacetime scale.  
However, it affects the spacetime distributions of collimated fireball properties, e.g., energy density, particularly approaching horizons. 
In this study, we attempt to obtain the approximate results and 
qualitative estimations of the total collimated fireball thermal energy, 
size and other quantities averaged over the entire collimated fireball volume and formation time at the end of collapse.
We also consider that such approximate results should indicate a lower bound of counterpart energetic quantities obtained in the GR framework, since the GR gravitational binding energy (potential) is larger than the Newtonian one in the gravitational collapse process from infinity to the horizon $2M$. The numerical approach based on the GR framework is necessary for quantitative studies of collimated fireball formation from gravitational collapses of massive rotating stellar and binary systems. 
It is fundamental for the fireball's inner part dynamics near the horizon. 
Nevertheless, due to the complex physical issues of GRB progenitors, we first figure out 
what macroscopic and microscopic processes are dominant and crucial for producing and collimating fireballs.
Therefore, it deserves preliminary analysis using the analytical approach based on the Newtonian approximation in the slow-rotating limit. The GR-based numerical simulation should not qualitatively change the general properties of physical effects and trends obtained in the Newtonian approximation.

In the Newtonian approximation, we describe the process by the continuity equation for the baryon number conservation, Euler's equation for energy-momentum conservation, and Poisson's equation for 
gravitational potential $\phi$ are,
\begin{eqnarray}
\frac{\partial \rho}{\partial t} +{\bf \nabla}\cdot(\rho \bf {u})&=& 0, \label{con}\\
   \Big(\frac{\partial {\bf u}}{\partial t}\Big)+ {\bf \nabla}\Big(\frac{1}{2}|{\bf u}|^2\Big) + (\nabla\times {\bf u})\times {\bf u} + \nabla h 
+ {\bf \nabla}\phi + {\bf \nabla}\phi_c &=& 0, \label{eulerT}\\
  \nabla^2 \phi -4\pi G \rho &=& 0,\label{potenT}
\end{eqnarray}  
where baryon fluid is described by density $\rho$ and velocity $\bf {u}$.
Equation of State (\ref{eos}) yields the baryon heat function $h=H/\rho$ 
($H$ for Enthalpy)
\begin{eqnarray}
\nabla h = \nabla p/\rho;  \quad h = \int \nabla p/\rho = \frac{\kappa \gamma}{\gamma -1}\rho^{\gamma -1}.
\label{heat} 
\end{eqnarray}
We formally add the gradient of centrifugal potential $\phi_c$ given by  ${\bf \nabla}\phi_c=j{\bf \nabla}\Omega$, 
where $j$ and $\Omega$ are angular momentum and velocity, respectively.
These equations reduce to their counterparts in the hydrodynamic equilibrium case, describing the core's state when the dynamic effects of collapse are not present.

\subsection{Static equilibrium configuration and rotation law}


We start with the established equilibrium conditions and rotation laws to develop a simple analytical model for a rigidly rotating stellar core undergoing gravitational homologous collapse. In hydrodynamic equilibrium, the integrability condition of Euler's equation (\ref{eulerT}) necessitates that the angular momentum $j$ depends only on angular velocity $\Omega$. We adopt the rotation law from Refs.~\cite{1989MNRAS.237..355K,Stergioulas_1998, 2002A&A...388..917D},
\begin{eqnarray}
j=A^2(\Omega_c-\Omega), 
\label{angula} 
\end{eqnarray}
where $\Omega_c$ is the value at the centre of the coordinate.  The rotation parameter $A$ is a positive constant (length scale), which implicitly depends on $\Omega_c$. 

In the Newtonian approximation, the rotation law simplifies to 
\begin{eqnarray}
\Omega &\approx & \Omega_c/(1+x^2/A^2),\label{angulb0}
\end{eqnarray}
where $x=R\sin\theta$ and $z=R\cos\theta$ are the cylindrical coordinates, $R$ is the radial coordinate, and $\theta$ is the zenith (polar) angle. The rotating axis is along the $\hat z$ direction. The system is axially symmetric, and we omit the coordinate $y$ in formulae. The centrifugal potential is given by 
\begin{eqnarray}
\phi_c &=& \int jd\Omega\approx A^2\Omega(\Omega_c - \Omega/2),
\label{angulb} 
\end{eqnarray}
and $d\phi_c=jd\Omega$.

We focus on the limit of constant angular velocity, namely a rigid rotating core. 
In this limit, 
\begin{eqnarray}
\Omega \rightarrow \Omega_c, \quad x^2/A^2 \ll 1, \label{slimit0}
\end{eqnarray}
and up to the leading order $\mathcal{O}(x^2/A^2)$, the rotation law becomes
\begin{eqnarray}
j \approx x^2 \Omega_c, \quad \phi_c \approx \frac{1}{2}x^2 \Omega^2_c\approx \frac{1}{2}  j \Omega_c.\label{slimit1}
\end{eqnarray}
Here, we drop the constant in Eq.~(\ref{angulb}) by ensuring the centrifugal potential 
to zero $\phi_c=0$ at the rotating axis $x=0$. We note that $hdm$ and $\phi dm$ are elementary heat and 
gravitational binding energies, $jdm$  and $\phi_cdm$ are elementary angular momentum and rotation (kinetic) energy for a baryon fluid element of mass $dm=\rho dV$ inside a rotating core.

In the static equilibrium case, integrating Euler's equation (\ref{eulerT}) yields
\begin{equation}
    h+\phi+\phi_c={\rm constant},
    \label{euler1Teq}
\end{equation}
which is uniform across the space. The constant on the right-hand side relates to the material binding energy. 
To generalize this to a model of a rigidly and slowly rotating stellar core undergoing homologous collapse, we need to consider the density, velocity, and potential changes during the collapse. 

\subsection{Homologous collapse of rigid rotating core}

To analyze the homologous collapse of a nearly spherically symmetric, rigidly and slowly rotating stellar baryon core, we follow the approach of Goldreich and Weber \cite{Goldreich1980} in the rotating frame. 
We assume the baryon fluid flow is vorticity-free, meaning $(\nabla \times \mathbf{u}) = 0$. The velocity ${\bf u}=\nabla v$
can thus be expressed as a gradient of a scalar function $v$:
\begin{eqnarray}
{\bf u}=\nabla v,\quad v=(1/2)(\dot a/a)R^2,\quad {\bf u}=(\dot{a}/a) {\bf R}=\dot{a}{\bf r}.\label{ufiled}
\end{eqnarray}
Here, $a(t)$ is a length scale function, and 
${\bf R}$ is the radial vector in the physical coordinates.
The core density $\rho$ and sound velocity $v_s$ (\ref{soundv}) in the core can be expressed in terms of a homologous function $f(r)$,
\begin{equation}
    \rho=\rho_{c}(t)f^3(r),
		\quad 
		v_s^2=\gamma p/\rho = (v_s^c(t))^2 f^{3(\gamma -1)}(r),
		\label{den1}
\end{equation}
where $r$ is
the dimensionless radius
\begin{equation}
r=R/a(t),\quad {\bf r}={\bf R}/a(t).	
\label{dimr}
\end{equation}
Here $\rho_c(t)$ is the central density, and $v_s^c(t)$ is the central sound velocity, given by $(v_s^c)^2\equiv\gamma p_c/\rho_c$. 
The length scale function $a(t)$ relates to the central density and sound velocity by:
\begin{equation}
   a(t)=v_s^c/(\gamma\pi G \rho_{c})^{1/2}.
		\label{asound}
\end{equation}
Given the functional forms of $\rho$ and ${\bf u}$, the continuity equation (\ref{con}) reduces to the trivial relation $\dot{f}=0$, indicating that the homologous profile $f(r)$ is time-independent.

The integrated Euler equation (\ref{eulerT}) yields
\begin{equation}
    \frac{\partial\upsilon}{\partial t}+\frac{1}{2}|{\bf \nabla} \upsilon|^2+h+\phi+\phi_c= {\rm constant},
    \label{euler1T}
\end{equation} 
which reduces to the equation (\ref{euler1Teq}) in the static equilibrium case $\dot a=0$ and $v=0$. 

By integrating these equations, the evolution of the collapsing core can be numerically simulated or approximately analyzed, incorporating the effects of rotation and centrifugal potential. 

\subsection{Spherical and homologous solution in space}\label{shs}

Neglecting $\phi_c$ and using the ansatz $v=(1/2)(\dot a/a)R^2$, Euler's equation (\ref{euler1T}) is separable in spatial and temporal variables,
\begin{eqnarray}
&&h+\phi=\frac{(v_s^c)^2}{\gamma-1}\frac{\lambda}{6} r^2,
\label{equi}\\
&&a^{\frac{\gamma}{2-\gamma}}\ddot a = -\frac{\lambda}{6} \frac{2\gamma}{\gamma-1}\left(\frac{\kappa^{\frac{1}{\gamma-1}}}{\pi G}\right)^{\frac{\gamma-1}{2-\gamma}}.
	\label{at}
\end{eqnarray}
where the positive eigenvalue $\lambda$ relates to the material binding energy of the baryon fluid per energy density 
$\rho$ \cite{Goldreich1980} or external pressure acting on the core surface. Namely, $\lambda dm$ relates to an elementary material binding energy for a baryon fluid
element of mass $dm$ around the nuclear density. 
Equation (\ref{at}) describes the time evolution of the length scale function $a(t)$ in the collapsing core.
The time-independent homologous spatial solution to Euler’s equation (\ref{equi}) is 
\begin{eqnarray}
h=\frac{(v_s^c)^2}{\gamma-1}f^{3(\gamma-1)},\quad \phi
=\frac{(v_s^c)^2}{\gamma -1}\left(\frac{\lambda}{6}~ r^2-f^{3(\gamma -1)}\right). 
\label{psi}
\end{eqnarray} 
Poisson's equation (\ref{potenT}) yields a second-order differential equation, 
\begin{equation}
    \frac{1}{r^2}\frac{d}{dr}\left(r^2 \frac{d f^{3(\gamma -1)}}{dr}\right)  
		+\frac{4(\gamma-1)}{\gamma}f^3=\lambda, 
		\label{e-profile}
\end{equation}
the boundary conditions are $f'(0)=0$ and $f(0)=1$. 

The mean thermal index $\gamma$ in the simple polytropic EoS (\ref{eos}), depending on the microscopic physics of material density and interactions, 
represents the average effect of repulsive heat energy pressure against the attractive gravity force.
Small $\gamma$ values for small heat pressure imply that stellar cores would be unstable and undergo gravitational collapse. 
Goldreich and Weber \cite{Goldreich1980} 
show that such EoS (\ref{eos}) and $\gamma=4/3$ 
provide a reasonable approximation to collapsing
stellar cores during the early phase before reaching nuclear density. In this simplified model, they obtain the homologous collapse solution of the onset value $\gamma=4/3$, below which massive cores undergo gravitational instability 
and collapse.

Following the same approach, we obtain \cite{Xue_2021} homologous collapse solutions for 
general values $1< \gamma < 4/3$, studying how spherical fireballs form in gravitational collapses. However, the realistic situations of gravitational collapses of massive cores approaching nuclear density are much more complex due to nucleon interactions and possible phase transitions at different densities. 
These are beyond the scope of the present study. 
In this article, we approximately consider EoS (\ref{eos}) and 
the mean thermal index $\gamma\gtrsim 1$ as an effective $\mathcal{O}(1)$ free parameter relevant to how much the gravitational 
binding energy gains for forming fireballs. 
The larger (smaller) $\gamma$ value is, the smaller (larger) gravitational binding energy gains. We select three $\gamma$ values (\ref{para}) for illustration.  

\begin{figure}
\centering
\includegraphics[width=3.0in]{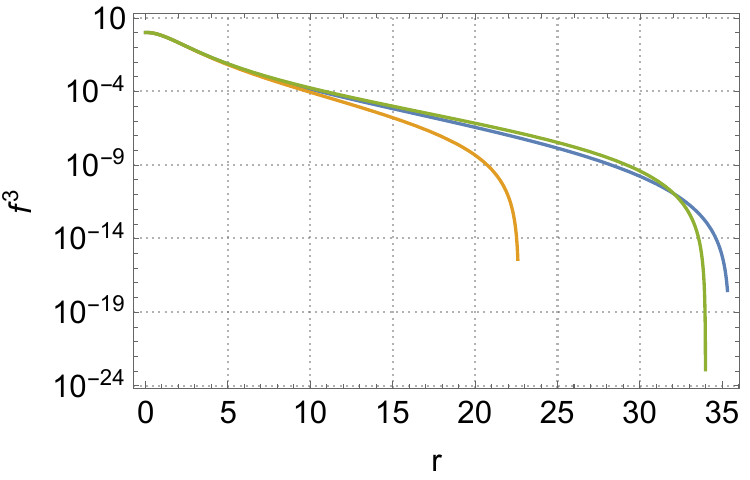}
\caption{Colours \orange{orange}, \blue{blue} and 
\green{green} on lines for three different eigenvalues selected in the list (\ref{para}). We show numerical solutions to the eigenvalue problem of Eqs.~(\ref{equi}) and (\ref{at}). The homologous density profile $f^3(r)$ (\ref{den1}) is independent of time. We plot it
as a function of dimensionless radius $r=R/a$. It vanishes $f^3(r_s)=0$ at the core outer radius $r_s$, see values in the list (\ref{para}). 
The increase of the core density $\rho=\rho_c(t)f^3$ (\ref{den1}) 
in homologous collapse is due to the increase of the centre density $\rho_c(t)$ via the decrease of the length scale function $a(t)$ (\ref{asound}) and Fig.~\ref{fa2}.}
\label{fa1}
\end{figure}

As shown in Fig.~\ref{fa1}, three solutions of homologous density profile $f^3(r)$ 
correspond to the selected values $(\gamma,\lambda)$:
\begin{equation}
\begin{tabular}{cccc}
Solutions & $\gamma$ &$\lambda$&$r_s$\cr
\hline
\hline
\orange{orange} & $1.24$& $1.0\times 10^{-4}$ & $23$\cr
\blue{blue} & $1.23$& $8.0\times 10^{-5}$ & $36$\cr
\green{green}& $1.225$& $8.0\times 10^{-6}$ & $34$\cr
\hline
\hline
\end{tabular}
\label{para}
\end{equation}\\
The homologous profile $f(r)$ becomes tangent to $f=0$ at the core outer radius $r_s$, namely 
$f'(r_s)\approx 0$. The core homologous profile $f(r)$ and outer radius $r_s$ do not depend on time. The core outer radius $r_s$ decreases (increases) as the material binding energy $\lambda$ value increases (decreases). 

In our analysis, we approximately decouple the centrifugal potential $\phi_c$ from the Euler equation (\ref{euler1T}), because 
the rotation energy is much smaller than the heat and gravitational potential energies. Namely, we neglect the back-reaction dynamics of the centrifugal potential $\phi_c$ in homologous gravitational collapse. The gravitational potential $\phi$ and heat function $h$ are balanced with the material binding energy $\lambda$ term in Eq.~(\ref{equi}). The homologously collapsing cores remain spherically symmetric without high moments.

The hydrodynamics of slow-rotating core collapse up 
until the core bounce can be understood in this relatively simple analytic model. 
This model provides a framework for understanding the centrifugal potential of a slowly and rigidly rotating core collapse until the core bounces, laying the groundwork for more detailed numerical simulations or further analytical refinements.

\begin{figure}
\centering
\includegraphics[width=3.0in]{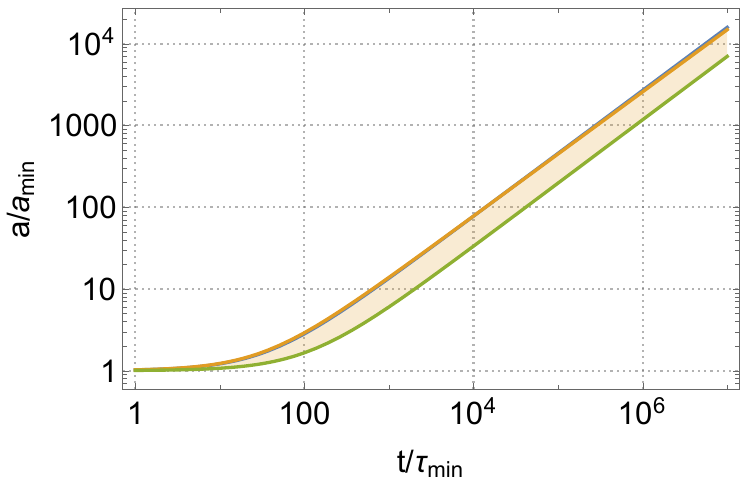}
\caption{Corresponding to the density profile $f^3(r)$ in Fig.~\ref{fa1}, the length scale  $a(t)/a_{\rm min}$ is plotted as 
in terms of $t/\tau_{\rm min}$, using units $a_{\rm min}$ (\ref{amin}) and $\tau_{\rm min}$ (\ref{taumin}). 
Note that the initial time $t_{\rm initial}\sim 10^6\tau_{\rm min}\sim {\mathcal O}(1)$ second, given by the initial conditions (\ref{inrho}). 
Colours \orange{orange}, \blue{blue} and 
\green{green} on lines for three different eigenvalues indicated in the list (\ref{para}). The shadow in between these colour lines indicates other possible physical solutions. 
}\label{fa2}
\end{figure}

\subsection{Initial central density and length scale function in time}\label{shst}

In homologous collapses, the baryon core centre density $\rho_c(t)$ increases as the length scale $a(t)$ (\ref{asound}) decreases. 
The temporal equation (\ref{at}) governs how the scale length $a(t)$ monotonically decreases in time. 
At the beginning of homologous collapses, i.e., $t_{\rm initial}$, the initial length scales and centre density $\rho^{\rm in}_c\equiv \rho_c(t_{\rm initial})$ are undetermined, varying from one homologous collapsing stellar system to another. In solving differential equation (\ref{at}), 
we chose the second constant of time integration by using boundary conditions at the end $t_{\rm final}$ of homologous collapses,  which give 
the minimal length scale $a_{\rm min}$ and the time scale
$\tau_{\rm min}$. We use the $a_{\rm min}$ ``sound horizon'' and $\tau_{\rm min}$ as the basic units for the temporal variables $a(t)$ and $t$. In such a way, the solution is written in a compact form, 
\begin{eqnarray}
	\frac{a(t)}{a_{\rm min}} &=& \left\{
1+ \Big[\frac{\lambda\gamma}{3(2-\gamma)(\gamma-1)^2}\Big]^{\frac{1}{2}}\Big(\frac{t}{\tau_{\rm min}}\Big)\right\}^{2-\gamma},
		\label{af}
\end{eqnarray}
which describes a monotonically collapsing 
process inverse time from the initial time $t_{\rm intial}\not=0$ to the final time $t_{\rm final}=0$. 
The length scale function $a(t)$ decreases significantly, 
approaching the minimum,
\begin{eqnarray}
a_{\rm min}&=& (3\gamma)^{\frac{2-\gamma}{2(\gamma-1)}}\Big(\frac{\kappa^{\frac{1}{\gamma-1}}}{\pi G}\Big)^{1/2}=3.06\times 10^{5} (3\gamma)^{-1/2} (\frac{\rho_c^{\rm max}}{\rho_0})^{-1/2}  ({\rm cm}).
\label{amin}\\
\tau_{\rm min} &\equiv& \Big(\frac{1}{\pi G \rho_c^{\rm max}}\Big)^{1/2}=1.02\times 10^{-5}(\frac{\rho_c^{\rm max}}{\rho_0})^{-1/2}  ({\rm sec}),
\label{taumin}
\end{eqnarray}
and $\tau_{\rm min}=(3\gamma)^{1/2}a_{\rm min}$. They give the basic macroscopic length and time scales of the homologous collapse.
The centre density $\rho_c$ and sound velocity $v_s^c$ increase in time,   
\begin{equation}
\rho_c=\rho_c^{\rm max} \Big(\frac{a}{a_{\rm min}}\Big)^{-\frac{2}{2-\gamma}};  \quad (v_s^c)^2= \frac{1}{3}\Big(\frac{a}{a_{\rm min}}\Big)^{-2\frac{\gamma-1}{2-\gamma}}=\frac{1}{3}\Big(\frac{\rho_c}{\rho_c^{\rm max}}\Big)^{\gamma-1},
\label{soundc}
\end{equation}
approaching their maximal values $\rho^{\rm max}_c=(3\gamma\kappa)^{-\frac{1}{\gamma-1}}$ and $1/\sqrt{3}$. 
The maximal centre density $\rho^{\rm max}_c=(3\gamma\kappa)^{-\frac{1}{\gamma-1}}$ is about $10$ times the nuclear saturation energy density 
$\rho_0\approx 2.4\times 10^{35} {\rm ergs}/{\rm cm}^3$, corresponding to the number density  
$n_0\approx 1.6\times 10^{38}/{\rm cm}^3$.  
The reason for choosing the maximal baryon core central density $\rho^{\rm max}_c\approx 10 \rho_0$ in this article and Ref.~\cite{Xue_2021} is the following. We consider a massive stellar baryon core at nuclear saturation 
density $\rho_0$, about the density of neutron stars. The baryon 
core conserves its total mass $M$ and continuously collapses to a compact object of size about $2GM$ of the black hole horizon. 
In this case, its average density increases and becomes
about $5\sim 8$ times the nuclear saturation energy density.

Instead of using collapsing time $t$, we choose the core central density $\rho_c(t)$ as the primary variable to describe the homologous gravitational collapse of a stellar core. This choice is physically intuitive and aligns with the key stages of the collapse, characterized by changes in the core central density. The initial value $\rho_c^{\rm in}=\rho_c(t_{\rm initial})$ depends on the homologous core mass $M$ and radius $r_s$. From Eqs.~(\ref{af}-\ref{soundc}), we obtain
\begin{eqnarray}
    \rho^{\rm in}_c\sim 10^{-6}\rho^{\rm max}_c 
    \label{inrho}
\end{eqnarray}
for $t_{\rm initial}\sim 10^6~ \tau_{\rm min}\sim {\mathcal O}(1)$ seconds. 
The final value $\rho_c^{\rm fi}=\rho_c(t_{\rm final})$ is the maximal value $\rho_c^{\rm max}$. The core undergoes a homologous collapsing process: starting from the initial time $t_{\rm initial}\sim {\mathcal O}(1)$ seconds and initial 
core centre density $\rho^{\rm in}_c$, ending at the final time $t_{\rm final}=0$ and final core centre density is $\rho^{\rm max}_c\approx 10 \rho_0$. The duration of homologous collapses $\Delta t=|t_{\rm final}-t_{\rm initial}|$ is about ${\mathcal O}(1)$ seconds.

\subsection{The rotation law of homologous collapse core}

We are now in the position of postulating the rotation law for a rigidly and slowly rotating stellar core, which undergoes homologous gravitational collapse. Since the stellar rotating scale $\Omega^{-1}$ and collapsing timescales $\tau_{\rm grav}=R/\dot R$ are much larger than microscopic interaction and thermalization timescales, we approximately consider the gravitational collapse of the rotating stellar core to be an adiabatic process. Namely, at each collapse step at time $t$, the stellar core is approximately in an equilibrium state as if the collapsing stellar core adiabatically evolves in a time sequence of equilibrium states during gravitational collapse. 

The rigid core angular velocity $\Omega_c(t)$ monotonically increases in time, as gravitationally homologous collapses proceed. We generalize the equilibrium rotation law (\ref{slimit1}) to
\begin{equation}
  j(t)=r^2\sin^2\theta a^2(t)\Omega_c(t);\quad  \phi_c(t)=\frac{[v^c_s(t)]^2}{2}r^2\sin^2\theta a^2(t)\Omega^2_c(t),
    \label{centrifugal}
\end{equation}
and $\phi_c(t)=[v^c_s(t)]^2j(t)\Omega_c(t)/2$ in the rotating frame. We parameterize the time-varying 
centrifugal potential $\phi_c(t)$ proportional to $[v^c_s(t)]^2$, 
analogously to the gravitational potential $\phi$ and heat function $h$ (\ref{psi}) in this semi-analytical model.

We will show soon that the rotating parameter $a^2(t)\Omega_c(t)
$ is time-independent, based on total core mass and angular momentum conservation.

\subsection{Total core mass and angular momentum conservations}

Using Eq.~(\ref{e-profile}) and outer boundary condition
$f'(r_s)\approx 0$ of the homologously collapsing core, and integrating $\int dm=\int dV \rho$ over an entire volume, we calculate its total core mass
\begin{eqnarray}
M = \int d^3R \rho &=& a^3(t)\rho_c(t)4\pi \int_0^{r_s} r^2 dr f^3 \\ \nonumber
&=& a^3(t)\rho_c(t)\frac{\lambda\gamma}{4(\gamma-1)}\frac{4\pi}{3}r^3_s,
\label{totalm}
\end{eqnarray}
and total core angular momentum
\begin{eqnarray}
J = \int d^3R \rho j &=& a^5(t)\rho_c(t)\Omega_c(t)\frac{8\pi }{3}\int_0^{r_s} r^4 dr f^3 \\ \nonumber
&\approx & a^5(t)\rho_c(t)\Omega_c(t)\frac{\lambda\gamma}{4(\gamma-1)}\frac{8\pi}{15}r^5_s,
\label{totall}
\end{eqnarray}
along the $\hat z$ direction in the rest frame.
The total core mass and angular momentum conservation lead to the constants \footnote{$a^3(t)\rho_c(t)=(\kappa/\pi G)^{3/2}$ for $\gamma=4/3$ 
} 
\begin{eqnarray}
a^3(t)\rho_c(t)= {\rm const.}, \quad a^2(t)\Omega_c(t)= {\rm const.},
\label{rlm}
\end{eqnarray}
in the homologous collapsing process from the initial core density $\rho_c^{\rm in}$ to the final core density $\rho_c^{\rm fi}=\rho_c^{\rm max}$. 
Equations (\ref{totalm}) and (\ref{totall}) 
give the ratio of total core mass and angular momentum
\begin{eqnarray}
J/M \approx (2/5) a^2(t)\Omega_c(t)r^2_s.\label{jmr}
\end{eqnarray} 
which is constant in time. This means the initial values from the beginning of collapse determine $J/M$ values. The larger $J/M$ values are, the larger rotating angular velocities $\Omega_c(t)$ are, and the $a^2\Omega^2_c$ values are large at the end of homologous collapses. The ratio $J/M$ is a physically sensible parameter not only for massive rotating core collapses, but also for massive rotating binary coalescence. Therefore, we will use the ratio $J/M$ to characterize rotating systems that undergo homologous collapses. 

\subsection{Slowly rotating approximation}

Analogously, using the solution of gravitational potential $\phi$ (\ref{psi}) we calculate the total core gravitational binding energy 
\begin{eqnarray}
U &=& \frac{1}{2}\int d^3R \rho \phi\nonumber\\
&=& a^3(t)\rho_c(t)\frac{[v^c_s(t)]^2(2\pi)}{(\gamma-1)} \int_0^{r_s} r^2 dr f^3\big[\frac{\lambda}{6}r^2-f^{3(\gamma-1)}\big] \nonumber\\
&=& M \frac{\lambda}{20}r_s^2\frac{[v^c_s(t)]^2}{(\gamma-1)}-M\frac{24}{r_s^3}\frac{[v^c_s(t)]^2}{\lambda \gamma}\int_0^{r_s} r^2dr f^{3\gamma} \nonumber\\
&\approx& -2 M\frac{[v^c_s(t)]^2}{3(\gamma-1)}.
 \label{totalu} 
\end{eqnarray}
Using the centrifugal potential (\ref{centrifugal}) $\phi_c$, 
we calculate the total core rotation energy as
\begin{eqnarray}
U_c = \int d^3R \rho \phi_c =\frac{[v^c_s(t)]^2 \Omega_c(t)}{2}\int d^3R \rho j= \frac{[v^c_s(t)]^2 \Omega_c(t)}{2} J.
\label{totalr}
\end{eqnarray}
It shows that as $\Omega_c(t)$ and $[v^c_s(t)]^2$ increase in 
homologous collapses, the core rotation energy $U_c$ increases by gaining gravitational binding energy. The ratio of total rotation energy and gravitational binding energy is
\begin{eqnarray}
U_c/|U| \approx \frac{3}{2}(\gamma -1)\Omega_c(t)(J/M). 
\label{ejgur}
\end{eqnarray}
The rotation energy is much smaller than the gravitational binding energy when $\Omega_c(t)J/M \ll 1$ 
for slowly rotating cores in homologous collapses.

It must be emphasized that the theoretical framework is qualitatively valid only for $\Omega_c(t)J/M \ll 1$, since (i) we neglect the centrifugal potential $\phi_c$ in studying a homologous collapse in Secs.~\ref{shs} and \ref{shst}; (ii) we 
assume the rotation law (\ref{centrifugal}) in homologous collapses. 
Nevertheless, we will see how the axial-symmetrically repulsive $\phi_c$ potential 
affects the spherically symmetrically attractive potential $\phi$. The latter converts the gravitational binding energy into the 
photon-pair thermal energy of fireballs at the end of a homologous collapse \cite{Xue_2021}. We will then get an insight into
the axial-symmetric distribution of photon-pair thermal 
energy (collimated fireball) when core rotations are present.

We make a few more emphases to end this section. The analytically tractable rigid core assumption (\ref{slimit0}) is a poor approximation for collapsing systems with large angular momenta, particularly for fast rotating cores and binary neutron star mergers. The differential rotation law (angular-momentum distribution law), as a simple example (\ref{angulb0}) or other complex rotation laws, should play a more prominent role. In this case, the core rotation energy $U_c$ should dissipate to the internal energy $h$, the length scale function $a(t)$ Eqs.~(\ref{centrifugal}) 
and (\ref{photonT}) should be altered. The axial symmetry along the rotating axis still holds and we may need two length scale functions $a(t)$ and $b(t)$ for an ellipsoid, because the part near the rotating axis with smaller centrifugal potentials more easily undergoes gravitational collapses than the part far from the rotating axis with larger centrifugal potentials. The angular momentum of gravitational collapsing systems gets redistributed, possibly forming a disc-like 
configuration, see reviews, for example, \cite{Liu2017}. However, it has become a difficult issue to handle even with numerical simulations. Furthermore, ultra-relativistic jets along the direction of rotating axes carry away the angular momentum. This further favours an axially symmetric disc-like configuration with characteristic radiation patterns. 

Moreover, we have not considered external magnetic field energy budgets \cite{Chandrasekhar1953} in gravitational collapsing systems, apart from the role of magnetic fields in the QED and QCD cascade multiplicity processes of photon production. Nevertheless, using the simplest model of rigid rotation, we obtain 
the total mass and angular-momentum conservation laws (\ref{totalm}-\ref{jmr}) and the essential quantity 
$J/M$ (\ref{jmr}) that are valid for general systems, and qualitative effects and trends that shed some light on the long-standing problem and provide a physical insight into complex numerical simulations. Further investigations are needed in future work.

\section{Collimated fireball energetic and geometric properties}\label{psphere} 

Based on the {\it adiabatic} approximation discussed in Sec.~\ref{Homo}, we can use the ``{\it local}'' virial theorem
(\ref{dgravT})  
to obtain the baryon heat energy $dF$ (\ref{dgravP}) and temperature $T$ (\ref{dgravT}). As discussed in Sec.~\ref{hadron}, the baryon heat energy is transformed 
into the thermal energy of a photon-pair plasma through hadronic photon production, resulting in a photon temperature (\ref{Tphoton}).

\subsection{Spherical fireball temperature of photons
and $e^+e^-$ pairs}

\begin{figure}[t]
\includegraphics[width=3.0in]{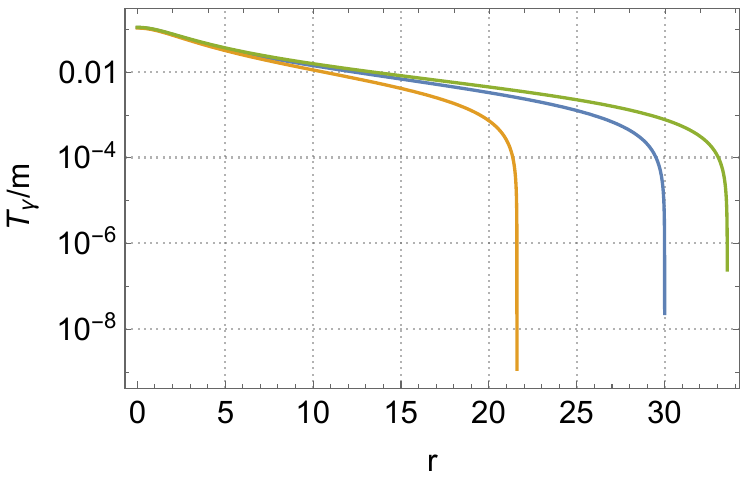}\includegraphics[width=3.0in]{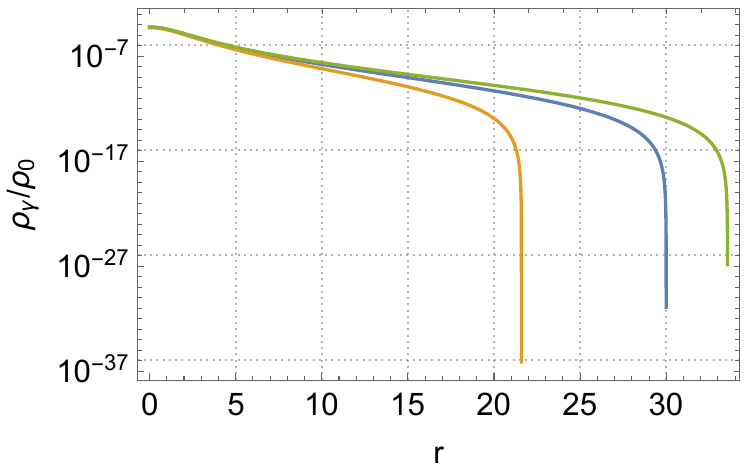}
\caption{Colours \orange{orange}, \blue{blue} and 
\green{green} on line for three eigenvalues (\ref{para}). Using results (\ref{dgravT1}) 
and (\ref{gammaEd}) for maximal sound velocity $v_s^c=1/\sqrt{3}$, we plot the homologous profiles of temperature $T_\gamma/m$ (left) and energy density $\rho_\gamma/\rho_0$ (right) 
as a function of dimensionless radius $r$.
Note that the typical hadron mass $m\approx 1$ GeV and 
nuclear saturation density $\rho_0\approx 2.4\times 10^{35} {\rm ergs}/{\rm cm}^3$. These homologous profiles do not depend on the time $t$. But their absolute values depend on time-dependent sound 
velocity $(v_s^c)^2(t)$, see Eqs.~(\ref{dgravT1}) and (\ref{gammaEd}).
The time evolutions of photon-pair sphere temperature $T_\gamma(r,t)/m$ (\ref{dgravT1}) and energy density  $\rho_\gamma(r,t)/\rho_0$ (\ref{gammaEd}) are given in the Appendix of Ref.~\cite{Xue_2021}.
}\label{ted}
\end{figure}

In non-rotating cases, centrifugal potential $\phi_c=0$, we obtain the baryon temperature $T$ by using the {\it local virial theorem} (\ref{dgravT}) and homologous solution (\ref{psi}).
The photon-pair temperature is given by 
$T_\gamma\approx 0.21 T$ (\ref{Tphoton}),
\begin{eqnarray} 
\frac{T_\gamma}{m} &\approx& 0.21 \frac{(v^c_s)^2}{6}\left\{\frac{1}{\gamma (\gamma -1)}
\left[(7\gamma -6)f^{3(\gamma-1)}(r) - \frac{\lambda}{6}\gamma r^2 \right]\right\}.\label{dgravT1}
\end{eqnarray}
The time-dependent centre sound velocity $v^c_s(t)$ comes from Eq.~(\ref{asound}). The temperatures $T=T(r,t)$ and 
$T_\gamma = T_\gamma(r,t)$ have the same 
spacetime configuration in homologous collapses. 
Moreover, we use 
photon productions (\ref{gammaN}) 
and (\ref{rhog}) by baryon collisions to approximately obtain the photon-pair spherical fireball energy and number densities
\begin{eqnarray}
\frac{\rho_\gamma}{\rho_0}&\approx & 1.29\times 
\frac{4\alpha\alpha_s}{3\pi}\Big(\frac{T_\gamma}{m}\Big)^4\ln\left( 1+ \frac{2.9}{4\pi\alpha_s}\right),
\label{gammaEd}\\
\frac{n_\gamma}{n_0} &\approx & 6.16 \times 
\frac{4\alpha\alpha_s}{3\pi}\Big(\frac{T_\gamma}{m}\Big)^3\ln\left( 1+ \frac{2.9}{4\pi\alpha_s}\right).
\label{gammaNd}
\end{eqnarray}

In Fig.~\ref{ted}, we plot the photon-pair temperature $T_\gamma$ and energy density $\rho_\gamma$ 
at the end of homologous collapses when the centre sound velocity $v^c_s(t)$ approaches $1/\sqrt{3}$.
The photon-pair temperature 
$T_\gamma$ reaches its maximum 
\begin{eqnarray}
\frac{T_\gamma^{\rm max}}{m} \approx 0.21  (v^c_s)^2 
\frac{(7\gamma -6)}{6\gamma (\gamma -1)},
\label{tmax}
\end{eqnarray} 
at 
photon-pair sphere centre $r=0$. We define the outer boundary $r_\gamma$ of the photon-pair sphere (spherical fireball) by $T(r_\gamma)=2m_e$, the energy threshold of electron-positron pair production. Compared $r_\gamma$ 
with the baryon core boundary $r_s$ in Fig.~\ref{fa1}, we find $r_\gamma <r_s$, namely the photon-pair sphere (fireball) is formed inside the baryon core, see also Fig.~\ref{sphere-jet1}.

{\it A posterior}, as a self-consistency check, 
we find that the baryon temperature $T$ and photon temperature $T_\gamma$ 
are much smaller than the baryon mass 
($T_\gamma \approx 0.21 T \ll m$). The condition is filled 
for applying the virial theorem (\ref{dgravP}). 

\begin{figure}[t]
\includegraphics[width=3.0in]{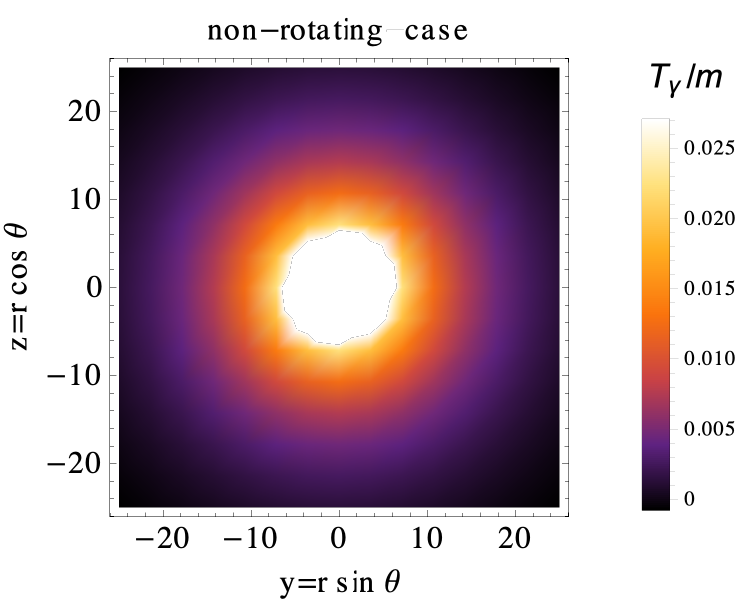}
\includegraphics[width=3.0in]{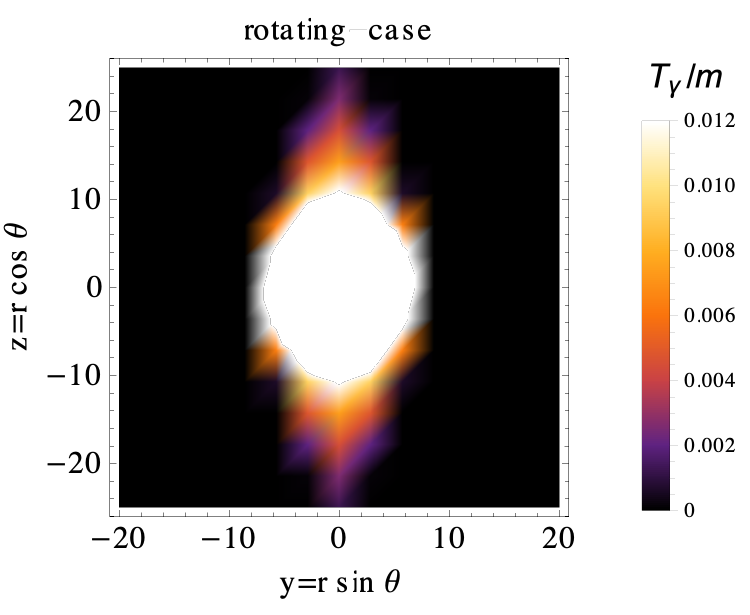}
\caption{Colours online. Using results (\ref{photonT}) for maximal sound velocity $(v_s^c)^2=1/3$, we plot the photon temperature $T_\gamma/m$ in the cylindrical coordinates
$y = r\sin\theta$ and $z = r\cos\theta$, where the $\theta$ is the zenith angle. The colourmap scale (linear) is indicated by the colour bar $T_\gamma/m$ in the left, where the temperature unit is of typical baryon mass $m\approx 1$ GeV. The parameters are the $\blue{blue}$ case parameters (\ref{para}), and the initial conditions are (\ref{inrho}). The left is a non-rotating case $a^2\Omega^2=0$. The right is a rotating case of $a^2\Omega^2=0.05$. The spherical fireball boundary $r_\gamma\lesssim 30$ 
and the collimated fireball boundary $r^j_\gamma(\theta)\leq r_\gamma$, 
see Fig.~\ref{sphere-jet1}. Both are smaller than the baryon core outer radius $r_s\approx 36$, see the list (\ref{para}) and blue curve in Fig.~\ref{fa1}. 
The baryon temperature $T/m$ distribution in the cylindrical coordinates
$y$ and $z$ is the same as the photon-pair temperature $T_\gamma=0.21 T$.}\label{sphere-jet}
\end{figure}

\subsection{Collimated fireball temperature of photons
and $e^+e^-$ pairs}

In rotating cases, centrifugal potential 
$\phi_c\not=0$ (\ref{centrifugal}), we obtain the baryon temperature $T$ by using the {\it local virial theorem} (\ref{dgravT}) and homologous solution (\ref{psi}). The photon-pair temperature is given by $T_\gamma\approx 0.21 T$ (\ref{Tphoton}), 
\begin{eqnarray} 
\frac{T_\gamma}{m} &\approx& 0.21 \frac{(v^c_s)^2}{6}\Big\{\frac{1}{\gamma (\gamma -1)}
\left[(7\gamma -6)f^{3(\gamma-1)} - \frac{\lambda}{6}\gamma r^2 \right]\nonumber\\
&-&\frac{1}{2}a^2\Omega^2_c r^2\sin^2\theta\Big\},\label{photonT}
\end{eqnarray}
at the end of homologous collapses, when $t=t_{\rm final}=\tau_{\rm min}$, minimal value $a(\tau_{\rm min})=a_{\rm min}$ (\ref{amin}) in Fig.~\ref{fa2}, maximal values $[v^c_s(\tau_{\rm min})]^2=(v^c_s)^2=1/3$, $\Omega_c(\tau_{\rm min})=\Omega_c$ and $a^2\Omega_c^2\equiv a^2(\tau_{\rm min})\Omega^2_c(\tau_{\rm min})$.
The first term inside the braces corresponds to the gravitational potential energy contribution to the temperature. 
The second term accounts for the rotational energy's influence, which decreases the photon-pair temperature due to the repulsive nature of the centrifugal potential $\phi_c$ acting against gravitational collapse. For $\theta\not=0$, 
the photon-pair temperature $T_\gamma$ becomes smaller, compared with non-rotating cases $\Omega_c=0$. Photon-pair fireballs are asymmetrically collimated along the rotation axes.

The expression for $T_\gamma$
highlights the interplay between the gravitational and centrifugal potentials in determining the thermal energy distribution of the photon-pair plasma during the homologous collapse of the rotating stellar core.
We describe the produced photon-pair plasma forms 
axial symmetric collimation in cylindrical coordinates. For illustration, we use the parameters
$(\gamma,\lambda_m)=(1.23,8\times 10^{-5})$, the \blue{blue} case in the list (\ref{para}), and $a^2\Omega^2_c=0.05$ for a slowly rotating core. 

\begin{figure}
\centering
\includegraphics[height=3.0in,width=3.0in]{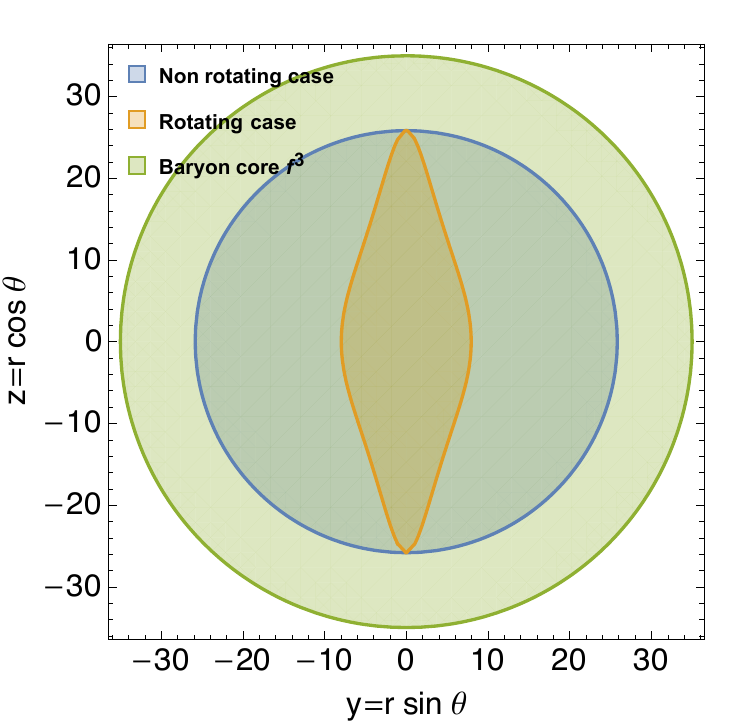}
\caption{Following the colourmap of Fig.~\ref{sphere-jet} for spherical and collimated fireballs, we compare the photon-pair fireball radius $r_\gamma$ and collimated fireball boundary $r^j_\gamma(\theta)$ (\ref{jetb}),
see Fig.~\ref{sphere-jet}, with the boundary $r_s\approx 36$ of the baryon core density $f^3(r)$ (\ref{den1}), see Fig.~\ref{fa1}. 
The photon-pair fireball radius $r_\gamma\lesssim 30$ and $R_\gamma=a_{\rm min}r_\gamma \sim (10^7-10^{10})$ cm. The photon-pair collimated 
fireball boundary $r^j_\gamma(\theta)$ is given in Fig.~\ref{sphere-jet2}. Note that the green zone and boundary indicate a baryon core, the blue (orange) zone and boundary indicate a spherical (collimated) fireball of photons and $e^+e^-$ pairs, and the boundary defines at the threshold $T_\gamma(r_\gamma)=2m_e$ ($T_\gamma[r^j_\gamma(\theta)]=2m_e$) respectively. 
The baryon temperature zone and boundary follow the photon-pair temperature zone and boundary for $T_\gamma=0.21 T$. 
These spatial distributions of fireballs and baryon core at the end of homologous collapses show $r_s>r_\gamma\geq r^j_\gamma(\theta)$. The collimated fireballs (orange zone) rotate along $\hat z$ 
axes with nontrivial angular momentum $J_\gamma$ (\ref{fireang}). They act as initial configurations for ultra-relativistic hydrodynamic evolution (early launching and later propagation): 
Spherical expansion or axial-symmetric jet evolution outwards, proceeding with internal and external shocks to produce GRBs.}
\label{sphere-jet1}
\end{figure}

\begin{figure}
\centering
\includegraphics[height=2.2in,width=2.7in]{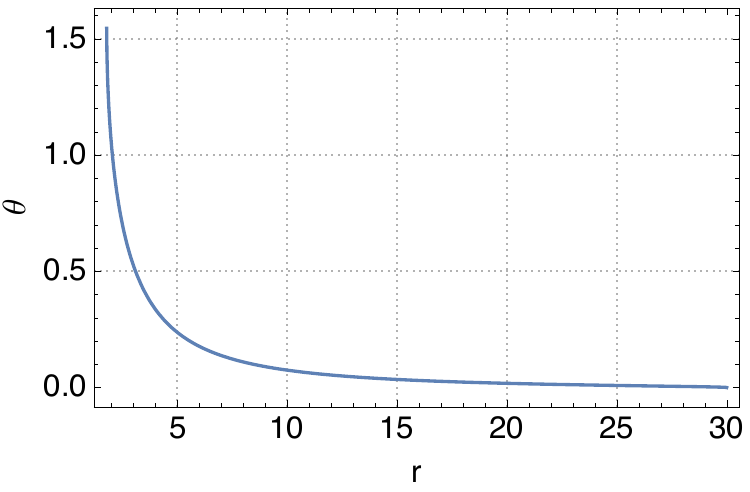}
\caption{The photon-pair collimated fireball boundary $r^j_\gamma(\theta)$ shows that (i) $\theta\rightarrow 0$ and $r^j_\gamma(\theta)$ approaches to the radius 
$r_\gamma\lesssim 30$ of the photon-pair fireball; (ii) $\theta\rightarrow \pi/2$ and $r_\gamma(\theta)$ approaches $r^j_\gamma(\pi/2)\approx 1$ of the photon-pair collimated fireball (the cylindrical collimation) width $R^j_\gamma
=a_{\rm min}r^j_\gamma(\pi/2) \sim (10^6-10^{8})$ cm, which is the maximal lateral size.
The collimated fireball angle (\ref{jeta}) 
$\theta_{\rm coll}\approx 1.8/30\approx 3.4^\circ$.
}\label{sphere-jet2}
\end{figure}

In Fig.~\ref{sphere-jet}, we plot the photon-pair
temperature $T_\gamma$ for 
the non-rotating case $\Omega_c=0$ and rotating cases $\Omega_c\not=0$. 
The maximal temperature is located at the centre $r=0$. 
The temperature $T_\gamma(r,\theta)=2m_e$ determines the photon-pair sphere fireball radius $r_\gamma < r_s$ or the collimated fireball boundary 
\begin{eqnarray}
r^j_\gamma (\theta) \leq r_\gamma,\quad  R^j_\gamma(\theta)=a(t)r^j_\gamma(\theta),
\label{jetb}
\end{eqnarray}
where the function $r^j_\gamma(\theta)$ is implicitly determined by
\begin{eqnarray}
\frac{2m_e}{m} &\approx&\frac{0.21 }{18}\Big\{\frac{1}{\gamma (\gamma -1)}
\left[(7\gamma -6)f^{3(\gamma-1)}(r^j_\gamma) - \frac{\lambda}{6}\gamma (r^j_\gamma)^2 \right]\nonumber\\
&-&\frac{1}{2}a^2\Omega^2_c (r^j_\gamma)^2\sin^2\theta\Big\}.
\label{jetc}
\end{eqnarray}
The photon-pair sphere volume $V_\gamma$ 
is larger than the photon-pair collimated volume $V^j_\gamma$, see Fig.~\ref{sphere-jet1}. In rotating cases, we plot the collimated boundary $r^j_\gamma(\theta)$ in Fig.~\ref{sphere-jet2}. 
As a self-consistency check, 
we find that the baryon temperature $T$ and photon temperature $T_\gamma$ are much smaller than the baryon mass 
($T_\gamma \approx 0.21 T \ll m$). It fills the condition for applying the virial theorem (\ref{dgravP}).

\subsection{Collimated fireball energy density and size profiles}

We use 
photon productions (\ref{gammaEd}) 
and (\ref{gammaNd}) by baryon collisions to approximately obtain the collimated fireball energy and number densities of photons and pairs
\begin{eqnarray}
\frac{\rho_\gamma}{\rho_0} \propto  \Big(\frac{T_\gamma}{m}\Big)^4,\quad
\frac{n_\gamma}{n_0} \propto \Big(\frac{T_\gamma}{m}\Big)^3,
\label{gammaNd1}
\end{eqnarray}
in terms of collimated fireball temperature $T_\gamma(r,\theta)$.
The results show that the fireball energy density $\rho_\gamma$ is in the range of $(10^{25}\sim 10^{30})~{\rm ergs}/{\rm cm}^3$, and
number density $n_\gamma$ is in the range of $(10^{37}\sim 10^{30})/{\rm cm}^3$. 

Using Eqs.~(\ref{af}), (\ref{amin}) and (\ref{soundc}), we obtain the photon-pair sphere radius
\begin{eqnarray}
R_\gamma &=& 3.06\times 10^{5}(3\gamma)^{-1/2}  r_\gamma \left(\frac{\rho_0}{\rho_c^{\rm max}}\right)^{\frac{\gamma-1}{2}} \left(\frac{\rho_0}{\rho^{\rm in}_c}\right)^{\frac{2-\gamma}{2}} ({\rm cm}),
\label{Rsize}
\end{eqnarray}
and the photon-pair collimation boundary
\begin{eqnarray}
R^j_\gamma(\theta) &=& 3.06\times 10^{5}(3\gamma)^{-1/2}  r_\gamma^j(\theta) \left(\frac{\rho_0}{\rho_c^{\rm max}}\right)^{\frac{\gamma-1}{2}} \left(\frac{\rho_0}{\rho^{\rm in}_c}\right)^{\frac{2-\gamma}{2}} ({\rm cm}).
\label{Rsizej}
\end{eqnarray}
They are expressed as a function of initial core centre density $\rho^{\rm in}_c$ when a homologous collapse initiates. 
The photon-pair sphere fireball 
size $R_\gamma\sim (10^{7}\sim 10^{10})~{\rm cm}$ and the photon-pair cylindrical collimation width $R^j_\gamma(\pi/2)\sim (10^{6}\sim 10^{9})~{\rm cm}$. 

Both sizes $R_\gamma$ and $R^j_\gamma$ are much larger than the photon mean-free path 
$\lambda_\gamma =(\sigma_\gamma n_\gamma)^{-1}$, namely $\sigma_\gamma n_\gamma R_\gamma\gg 1$ or $\sigma_\gamma n_\gamma R^j_\gamma\gg 1$. The
photon-pair spherical
and collimated fireballs are deeply opaque. 
On the other hand, we find in Fig.~\ref{sphere-jet} that the photon-pair spherical and collimated fireballs' 
temperature $T_\gamma$ can be well above the critical energy threshold 
$2m_e=1.02$ MeV of electron-positron ($e^+e^-$) pair production. 
The photon-pair spherical/collimated fireball
energy densities $\rho_\gamma$ can be well above the 
critical energy density $\rho_e=m_e^4=5.93\times 10^{-11}\rho_0$ 
of electron-positron pairs.
This means the photons 
are quickly thermalised to form an electron-position-photon plasma, as briefly  
discussed in Sec.~\ref{virial}. 

Up to now, we have studied in this simplified model the processes which the gravitational
binding energy is extracted and converted to a collimated 
and opaque fireball energy at the end
of gravitational stellar collapses. The effect of an external magnetic field in such a process needs further investigation. These studies of extracting gravitational binding energy in stellar collapses contrast with the processes \cite{1969NCimR...1..252P, Penrose1971,1977MNRAS.179..433B} of extracting rotational energy from spinning black holes.

\subsection{Collimated fireball energy and spectrum peak}

The total relativistic particle energy $E_\gamma$ and number $N_\gamma$ of the photon-pair collimated fireball can be obtained by numerically integrating the energy and number densities (\ref{gammaNd}) over the photon-pair collimated volume 
\begin{eqnarray}
V^j_\gamma=\int_{V^j_\gamma} d^3R=a^3\int_{V^j_\gamma} 2\pi r^2dr \sin\theta d\theta.
\label{Vsizej}
\end{eqnarray}
We also compute the total baryon core 
mass 
\begin{eqnarray}
M=\int_{\rm core} \rho d^3R=\rho_c a^3F_0,
\label{tmassj}
\end{eqnarray}
and number $N=M/m$, where $F_0\equiv \int_0^{r_s} 4\pi r^2 drf^3$. 

At the end of homologous collapse, when the central sound velocity $v^c_s=\sqrt{1/3}$, we 
define integration 
\begin{eqnarray}
Y_n =  
2\pi \int_{V^j_\gamma} \left(\frac{T_\gamma}{m}\right)^n r^2dr \sin\theta d\theta 
\label{Yn}
\end{eqnarray}
over photon-pair collimated volume $V^j_\gamma$. 
From Eq.~(\ref{gammaNd}),
we obtain the total energy and photon number of photon-pair collimated systems 
\begin{eqnarray}
\frac{E_\gamma}{M}
&\propto & \frac{Y_4}{F_0}\big(\frac{\rho_c^{\rm max}}{\rho_0}\big)^{4(1-\gamma)}
\big(\frac{\rho_0}{\rho^{\rm in}_c}\big)^{5-4\gamma},
\label{tote}\\ 
\frac{N_\gamma}{N}
&\propto&  \frac{Y_3}{F_0}\big(\frac{\rho_c^{\rm max}}{\rho_0}\big)^{3(1-\gamma)}
\big(\frac{\rho_0}{\rho^{\rm in}_c}\big)^{4-3\gamma}.
\label{totn}
\end{eqnarray}
In addition, we define the mean temperature of the photon-pair sphere/collimation as averaged 
value of $T_\gamma$ (\ref{photonT}) 
over photon-pair collimated volume $V^j_\gamma$, 
\begin{eqnarray}
\frac{\langle T_\gamma\rangle}{m}&\propto& \frac{Y_1}{Y_0}\left(\frac{\rho_0}{\rho_c^{\rm max}}\right)^{\gamma-1}\left(\frac{\rho^{\rm in}_c}{\rho_0}\right)^{\gamma-1},
\label{atem}
\end{eqnarray}
which represents the characteristic temperature of the photon-pair sphere/collimation. 
The photon-pair spectrum  
has a maximum at the peak energy $E_p \sim \langle T_\gamma\rangle$.  

In Eqs.~(\ref{tote}) and (\ref{atem}), we express total photon-pair energy $E_\gamma$ and spectral peak $E_p$ in terms of the initial 
core centre density 
$\rho^{\rm in}_c=\rho_c(t_{\rm initial})$ 
at the time $t_{\rm initial}$ when the core initiates homologous collapse. The final core centre density is about 10 times the nuclear density $\rho^{\rm max}_c\approx 10 \rho_0$. 
If the core mass $M$ and baryon number $N$ 
are fixed and conserved, the smaller initial core centre density $\rho^{\rm in}_c$ is, the more extended core mass density profile $\rho(r,t_{\rm initial})$ is, i.e., the larger $r_s$ is.
As a consequence, more gravitational energy can be gained in collapse 
and converted to the photon-pair sphere/collimation energy $E_\gamma$ and number $N_\gamma$. 
In contrast, these quantities are also functions of the material binding energy parameter $\lambda$, the averaged 
thermal index $\gamma$ and the rotation parameter ${\mathcal A}^2a^2\Omega_c^2\propto J/M$.
The rotation represents the repulsive reactions against gravitational collapse and 
energy gain. 

For gravitational collapses from the initial core density $\rho_c^{\rm in}\sim 10^{-6}\rho_0$ to the final one $\rho_c^{\rm max}\sim 10 \rho_0$, we obtain
\begin{eqnarray}
\frac{E_\gamma}{M}\sim (2\sim 5)\times 10^{-2},\quad \frac{E_p}{m}\sim (8\sim 10)\times 10^{-5}\nonumber\\
\frac{N_\gamma}{N},\sim (1\sim 3)\times 10,\quad R_\gamma\sim (5\sim 9)\times 10^8{\rm cm},
\label{enegetics}
\end{eqnarray}
where the variations are due to the ranges of the parameters $\lambda$, $\gamma$, and small values 
$a^2 \Omega_c^2 \propto\Omega_c (J/M)$. These results are consistent with the total energy and entropy budgets required to explain the cosmological origin of the GRB phenomena observed. 
We recall that in the Newtonian approximation, the variation of gravitational energy $GM^2/(2R)$ is $M/4$ for a core-collapse from infinity $R=\infty$ to $R=2GM$.

The realistic situations of baryon collisional relaxation and photon production in gravitational collapse and hydrodynamic processes are much more 
complicated. Among others, we mention a few questions about 
internal perturbations and shock waves formed due to the baryon matter EoS (\ref{eos}) and heat function variations, phase transition 
and material binding energy change as nuclear matter density variations, fluid plasma stability in external magnetic fields and general relativistic effects.  
We expect the reduction of collimated fireball energy and number densities, the softened energy spectra, and the prolonged collapsing time. These questions will be the subject of future studies.

\section{Collimated fireball angular momentum and relativistic jet}\label{jetf}

\subsection{Angular momentum conservation making an ultra-relativistic jet}

The collimated photon-pair fireball is neutral and free from the classical forces of large-scale electromagnetic fields since there are no macroscopic distributions
of electric charges and currents. In addition, the collimated fireball carries a nontrivial angular momentum $J_\gamma$ along the $\hat z$ direction in the rest frame, 
\begin{eqnarray}
J_\gamma/M = M^{-1}\int d^3R \rho_\gamma j, \quad J_\gamma/M\ll J/M,\quad J_\gamma\propto J
\label{fireang}
\end{eqnarray}
analogously to the total core angular momentum $J$ 
calculations (\ref{totalm}). The $jdm_\gamma=j\rho_\gamma dV$ is an elementary angular momentum of the photon-pair fluid element of energy $dm_\gamma = \rho_\gamma dV$ inside a collimated fireball. 
Since the fireball thermal temperature $T_\gamma$ 
is much smaller than the baryon mass scale $m$, $T_\gamma \ll m$ shown in Fig.~\ref{sphere-jet}, 
the collimated fireball energy density $\rho_\gamma\propto T_\gamma^4$ (\ref{gammaNd}) is much smaller than the baryon core density $\rho_\gamma\ll \rho$, therefore $J_\gamma\ll J$. Equations (\ref{totalm}) and (\ref{fireang}) show the collimated fireball angular momentum $J_\gamma$ is proportional to the 
angular momentum $J$ of massive rotating systems in gravitational collapses. 
The total angular momentum conservation in gravitational collapsing processes implies that collimated fireballs acquire angular momentum $J_\gamma$ from rotating systems in collapses.

The hydrodynamical outflows of the collimated fireballs due to the huge photon-pair pressure $p_\gamma \sim \rho_\gamma$ are ultra-relativistic. The hydrodynamical outflows of the collimated fireballs due to the extremely high photon-pair pressure $p_\gamma \sim \rho_\gamma$ are ultra-relativistic. The collimated fireball outflow carries away the energy $E_\gamma$ and angular momentum $J_\gamma$ from collapsed systems, engulfing the outer material 
and forming an ultra-relativistic ejecta interacting with stellar matter, as described in the standard GRB 
scenario. 
The novelty is that the collimated outflows and ejecta launch ultra-relativistic jets of nontrivial angular momenta $J_\gamma$ of collimated fireballs. 
In later propagation, ultra-relativistic jets maintain collimated structure due to the conservation of their nontrivial angular momentum $J_\gamma\not=0$, analogously to a gyroscope effect. Numerical/simulation-based studies are necessary for verifying such dynamics.
Analytical studies of ultra-relativistic jets launching and propagation, and their stability in magnetic fields, will be topics for separate articles.  
Notwithstanding, we present in the next section some qualitative discussions on the ultra-relativistic jet formation, early launching and later propagation.

In contrast, hydromagnetic flows from accretion discs transport away the rotational energy and angular momentum along large-scale toroidal magnetic fields, making a relativistic jet \cite{1982MNRAS.199..883B}.

\subsection{Collimated fireball angle $\theta_{\rm coll}$ relates to jet angle}\label{cangle}

As depicted in Figs.~\ref{sphere-jet} and \ref{sphere-jet1}, the collimated 
fireball of photon-pair plasma energy density $\rho_\gamma$ 
shows denser and wider at the base $(\theta\rightarrow\pi)$, while less dense and narrower at the top $(\theta\rightarrow 0)$. It is the consequence of the gravitational binding potential $\phi$ and centrifugal potential 
$\phi_c$ in the temperature profile $T_\gamma$ (\ref{photonT}) at the end of homologous collapses. The opaque collimated fireball 
pressure (density) $p_\gamma\sim \rho_\gamma\propto T_\gamma^4$ is no longer isotropic and its maximal value along 
$\hat z$-axis, decreasing from $R=0$ to $R=R_\gamma$. In the transverse directions perpendicular to $\hat z$, its value decreases to the collimated fireball boundary (orange thick line in Fig.~\ref{sphere-jet1}). Namely, most collimated fireball energy concentrates at the centre along $\hat z$. 

On the contrary, the collimated fireball's angular momentum $j\rho_\gamma dV$ 
vanishes along $\hat z$ direction ($\theta=0$) from 
$R=0$ to $R_\gamma$, and increases in the transverse 
directions ($\theta=\pi/2$). Namely, most fireballs' 
angular momentum is carried near the collimated fireball boundary. An outward thermal pressure in the 
fireball centre along the $\hat z$ direction and the conservation of angular momentum near the collimated fireball boundary possibly leads to an ultra-relativistic jet. 

In the phase of early jet launching, we argue that such a collimated fireball will self-similarly evolve to the normal jet profile of energy density denser and wider at the top $(\theta\rightarrow 0)$, less dense and narrower at the base $(\theta\rightarrow \pi)$. 
The reason is that
the highly dense and energetic part is deeply opaque 
and has a large thermal pressure $p_\gamma$ along the $\hat z$ direction, ultra-relativistically moving outward. As a result, denser and more energetic parts pile up on the front, moving at almost light speed, while dilute and less energetic parts leave behind a narrow tail. 
It is also the case for the collimated fireball outflow engulfing a small amount of material. 
This situation is analogous to the spherical case of 
ultra-relativistically hydrodynamic expansion: a spherical fireball evolves to a self-similar (homologous) spherical slab at the front moving at almost light speed, and a tail of dilute number and energy densities left behind, which are studied both analytically and numerically \cite{Ruffini1999, Ruffini2000}. 

We must mention the condition of a collimated fireball forming an outward-propagating jet. Figure \ref{sphere-jet1} shows that collimated fireballs form inside external baryon shells (green layers). 
Whether or not they can erupt out of external baryon shells requires the ratio $E_\gamma/M_{\rm ej}\gg 1$, see the 
references and discussions for Eq.~(\ref{gamma0}), 
where $E_\gamma$ is the total energy of collimated fireballs and $M_{\rm ej}$ is the ejecta mass contributing from the external baryon shell. The $M_{\rm ej}$ is much smaller than the total baryon core mass $M$, i.e., $M_{\rm ej}\ll M$, 
because Figure \ref{fa1} shows the baryon external shell mass density is very small for $r>10$ comparing the core centre density. While collimated fireball energies are 
about $E_\gamma/M \sim 10^{-2}$ (\ref{enegetics}). This implies the possibility of collimated fireballs erupting from baryon cores and forming ultra-relativistic ejecta with nontrivial angular momenta.

In the phase of later jet propagation, we argue that such a formed ultra-relativistic jet maintains its structure and dynamics due to the conservation of its jet angular momentum $J_\gamma$ along the jet propagating direction. The jet angle $\theta_{\rm jet}\approx dR_\perp/dR$ can be defined with its radius $R$ and lateral size $R_\perp$. However, one has to consider the ultra-relativistic jet interaction with ambient stellar media and its stability in external large-scale magnetic fields. These issues necessarily require numerical studies.

Nevertheless, we compare the photon-pair sphere and collimated boundaries in Fig.~\ref{sphere-jet1} and define the cylindrical collimated fireball angle $\theta_{\rm coll}$ as the ratio 
\begin{equation}
\theta_{\rm coll}\approx \frac{R^j_\gamma(\theta=\pi/2)}{R^j_\gamma(\theta=0)},\quad R^j_\gamma(\theta=0)=R_\gamma,
\label{jeta}
\end{equation}
where $R^j_\gamma(\theta=\pi/2)$ is the maximal lateral size of collimated fireballs and $R^j_\gamma(\theta=0)$ is the 
corresponding radius of the spherical fireball.
Such an angle $\theta_{\rm coll}$ characterizes the axially symmetric collimation of photon-pair fireball along the $\hat z$ direction before its ultra-relativistic hydrodynamic outflow. 

The collimated fireball angle $\theta_{\rm coll}$ (\ref{jeta}) differs from the 
ultra-relativistic outflow jet angle $\theta_{\rm jet}$ and the observational jet 
angle $\theta^*_{\rm jet}$ defined through the maximal lateral size of the equal arrival time surface in the structure and dynamics of GRB jets \cite{Granot:2006hf}. 
The collimated structure of the ultra-relativistic outflow should be maintained 
in jet launching and propagation due to the conservation of the angular momentum $J_\gamma$ along the jet direction. Therefore, we speculate that the GRB jet angles $\theta_{\rm jet}$ and $\theta^*_{\rm jet}$ should proportionally relate (or positively correlate) to 
the collimated fireball angle $\theta_{\rm coll}$,
\begin{equation}
\theta_{\rm coll}\sim \theta_{\rm jet}\sim \theta^*_{\rm jet}.
\label{jeta1}
\end{equation}
The smallness of the collimated fireball angle $\theta_{\rm coll}$ implies 
the smallness of the GRB jet angles $\theta_{\rm jet}$ and $\theta^*_{\rm jet}$.

We make some remarks to end this section. 
The collimated fireball energy $E_\gamma$ and ejector mass $M_{\rm ej}$ ratio $\Gamma_0\approx E_\gamma/M_{\rm ej}>1$ (\ref{Gamma0}) is one of the necessary conditions for a collimated fireball erupting out of external baryon shells (stellar envelope) as an outward collimated flow, accompanied by cocoon configurations \cite{Nakar2017}. However, to maintain a distinct ``thin'' jet structure in its ultra-relativistic propagation and penetrate through 
the stellar envelope, the collimated outflow should be energetic $E_\gamma/M_{\rm ej}\gg 1$ and carry large enough angular momentum $j_\gamma/ M$ (\ref{fireang}) to overcome small shearing forces (torques) from the surrounding material. It is the case particularly
for large $M$ and $J/M$  binary coalescence (short GRBs), producing small $\theta_{\rm coll}$  ``thin” jets, which carry away more angular momentum along the rotation axes, and interact with less massive $M_{\rm ej}$ stellar envelopes, resulting in weak cocoons, more disc-like configurations and strong gravitational wave signals. 

Otherwise, for small values $E_\gamma/M_{\rm ej}>1$ 
and $j_\gamma/ M$, outward collimated flows become relativistic shocked jets, due to dissipation of their energy and angular momentum to the surrounding material (Newtonian shocked stellar material), creating cocoon configurations. It is the case particularly for massive $M$ and small $J/M$ core collapses (long GRBs), producing large $\theta_{\rm coll}$ ``fat'' jets interacting with more massive $M_{\rm ej}$ stellar envelopes, resulting in strong cocoons, less disc-like configurations and weak gravitational wave signals.

The fractions of shocked jet and stellar material components in cocoons depend on how collimated fireballs transfer their energies $E_\gamma$ and 
angular momenta $j_\gamma$ to the stellar envelope materials.
The shocked jet angle $\theta_{\rm jet}$ should be smaller than the collimated fireball angle $\theta_{\rm coll}$,
because of angular momentum transfers. Less energetic collimated fireballs $E_\gamma/M_{\rm ej}\gtrsim 1$ 
become ``failed'' jets, creating hot spots of Newtonian shocked stellar materials. These cases should be the most probable fates of massive stellar core collapses, but their low-frequency signals are probably too faint for recent observations.

\section{Collimated fireball intrinsic correlations}\label{corre}

In this simplified model, we obtain the configurations of photon-pair collimated fireballs determined by 
\begin{eqnarray}
(J/M,\rho^{\rm in}_c/\rho_0, \rho^{\rm fi}_c/\rho_0,\gamma,\lambda),  
\label{fpara}
\end{eqnarray}
where the total mass and angular momentum ratio $J/M$  characterizes rotating systems in gravitational collapsing processes, the initial and final centre densities $\rho^{\rm in}_c/\rho_0$ and $\rho^{\rm fi}_c/\rho_0$ crucially relate to the gravitational binding energy gain, averaged thermal index $\gamma$ and material binding energy $\lambda$ represent material reactions in processes. 
The photon-pair collimated fireball energetic properties (\ref{Rsize},\ref{Rsizej}) and (\ref{tote},\ref{atem}) explicitly depend on the initial and final core central densities $\rho_c^{\rm in}$ and $\rho_c^{\rm fi}=\rho_c^{\rm max}\approx 10\rho_0$. They determine the gravitational binding energy gain and conversion to baryon kinetic energy and photon-pair thermal energy in gravitational collapses. These characteristics should also be held for gravitational collapses of non-spherical cores or binary mergers by introducing effective initial and final central densities $(\rho_c^{\rm in})_{\rm eff}$ and 
$(\rho_c^{\rm max})_{\rm eff}$ of the system. 
For appropriate configurations (\ref{fpara}), this scenario could also provide an efficient mechanism of photon-pair driven supernova explosion, and detailed investigations are required.

\subsection{Intrinsic correlations among collimated fireball quantities}

All collimated fireball quantities (\ref{Rsize},\ref{Rsizej}) and (\ref{tote},\ref{totn},\ref{atem}) are expressed as scaling laws of the initial core centre density $\rho_c^{\rm in}$ since it characterizes how much the gravitational binding energy gains when the systems of masses $M$ gravitationally collapse from $\rho_c^{\rm in}$ to final core centre density $\rho_c^{\rm fi}=\rho_c^{\rm max}\approx 10 \rho_0$.

By simplifying out the initial core centre density 
$\rho_c^{\rm in}$ 
among  
Eqs.~(\ref{Rsize},\ref{Rsizej}) and (\ref{tote},\ref{atem}), we obtained
theoretical correlations among the spectral peak $E_p$, total photon-pair energy $E_\gamma$ and peak luminosity $L_\gamma \propto c R_\gamma^2\langle T_\gamma\rangle^4\propto cE_\gamma/R_\gamma$ of the photon-pair sphere or collimation in a local rest frame,
\begin{eqnarray}
E_p \propto E_\gamma^\chi,\quad
E_p\propto 
L_\gamma ^{\frac{\chi}{\chi+1}},\quad
L_\gamma \propto E_\gamma^{1+\chi},  
\label{chile}
\end{eqnarray}
which depend on the unique parameter 
$\gamma\gtrsim 1$
and $\chi= \frac{\gamma-1}{4\gamma -5}\sim {\mathcal O}(1)$ \footnote{It is an inverse of the original $\chi-$definition \cite{Xue_2021}.}.

These collimated fireballs undergo ultra-relativistic expansion and interact with massive stellar material, forming ultra-relativistic ejecta whose Lorentz factors can be estimated as \cite{2004RvMP...76.1143P,zhang_2018, 2012grb..book.....K, Ruffini1999, Ruffini2000, Meszaros2000, Zhang2021}
\begin{equation}
\Gamma_0 \approx E_\gamma/M_{\rm ej}\label{Gamma0} 
\end{equation}
where $M_{\rm ej}$ are ejected masses. This article does not present studies of ultra-relativistic outflow of collimated fireballs. However, we attempt to explain the ejected masses $M_{\rm ej}$ in this simplified model, thus to have an estimation of Lorentz factors (\ref{Gamma0}).

At the end of homologous collapse, the homologous baryon core density 
$\rho$ (\ref{den1}) profile (Fig.~\ref{fa1}) and photon-pair collimated fireball profile (Fig.~\ref{sphere-jet}) show that the baryon core size $r_s$ is larger than photon-pair collimated fireball size $r_\gamma$, e.g.,  $r_s>r_\gamma$ (\ref{jetb}). 
This result implies that the photon-pair collimated fireball
expansion erupts and pushes/engulfs 
the outer part (tail) 
of collapsing baryon cores outward. Therefore, the outer part of the baryon core density contributes to the ejecta masses $M_{\rm ej}\ll  M$.  
We approximately assume the positive correlation between the ejecta mass $M_{\rm ej}$ and initial centre mass density $\rho^{\rm in}_c$, i.e., $M_{\rm ej}\propto (\rho^{\rm in}_c)^\beta$. The reason is that the homologous core density profile shows 
that total collapsing core masses $M$ are large, centre mass densities $\rho^{\rm in}_c\propto M$ are 
large and outer part masses $M_{\rm ej}\propto M$ are large. The positive parameter $\beta$ is of order unity, and we set $\beta=1$ for simplicity.

The total collimated fireball 
energy is $E_\gamma\propto (\rho^{\rm in}_c)^{(4\gamma-5)}$ (\ref{tote}). Therefore, 
the ejecta Lorentz factor is approximately given by
\begin{eqnarray}
\Gamma_0 \approx  E_\gamma/M_{\rm ej} \propto (\rho_c^{\rm in})^{(4\gamma-5)-1}. 
\label{gamma0}
\end{eqnarray}
The collimated fireball has size scale $R_\gamma \propto (\rho_c^{\rm in})^{\frac{\gamma-2}{2}}$ (\ref{Rsize}). Its time scale $\tau_\gamma=R_\gamma/c$ relates to an observational one $T_{90}$
by the approximate relation $T_{90} \approx  \tau_\gamma/(2\Gamma^2_0)$ \cite{Piran_1999}, 
\begin{eqnarray}
T_{90}&\approx & \tau_\gamma/(2\Gamma^2_0)\propto (\rho_c^{\rm in})^{-\frac{2-\gamma}{2}-2(4\gamma-5)+2}.\label{t90}    
\end{eqnarray}
As a result, we approximately express the ejecta Lorentz factor $\Gamma_0$ and GRB observational time scale $T_{90}$ as scaling
laws of the initial core centre density $\rho_c^{\rm in}$.

From the $\{\tau_\gamma, E_\gamma,\langle T_\gamma \rangle, L_\gamma\}$ universal $\chi$-correlations (Eqs.~(7.2-7.7) in Ref.~\cite{Xue_2021}), we obtain the $\{T_{90}, E_{iso}, E_{p}, L_{ iso}\}$ universal $\chi$-correlations: 
\begin{eqnarray}
E_{p} &\propto& E_{iso}^{^\chi}, 
\label{chi}\\
E_{p} &\propto& L_{iso}^{\frac{\chi}{3+\chi-\chi_{_\Gamma}}},
\label{chil}\\
L_{iso} &\propto& E_{iso}^{3+\chi-\chi_{_\Gamma}}.  
\label{chile1}
\end{eqnarray}
and
\begin{eqnarray}
E_{p} &\propto& T_{90}^{-\frac{2\chi}{3(1+\chi)-2\chi_{_\Gamma}}},\label{time1}\\
E_{iso} &\propto& T_{90}^{-\frac{2}{3(1+\chi)-2\chi_{_\Gamma}}},\label{time2}\\
L_{iso} &\propto& T_{90}^{-2\frac{(3+\chi)-\chi_{\Gamma}}{3(1+\chi)-2\chi_{_\Gamma}}},
\label{time3}
\end{eqnarray}
where 
$\chi_{_\Gamma}=2/(4\gamma-5)$, and
they can be recast as other correlations.
Based on the simplified model, we obtain the theoretical correlations (\ref{chi}), (\ref{chil}) and (\ref{chile}), 
anti-correlations (\ref{time1}), (\ref{time2}) and (\ref{time3})
in a unified framework.
They represent intrinsic properties of the photon-pair collimated fireballs and should be universal for all GRBs' progenitors. 

We cannot analytically obtain intrinsic correlations between the collimated fireball angle $\theta_{\rm coll}$ (\ref{jeta}), the ratio $J/M$ and other collimated fireball quantities $E_p^j$, $E_\gamma^j$, $L_\gamma^j$ and $R_\gamma^j$. However, in the 
next Sec.~\ref{spherevsjet}, we will try to obtain the numerical trends (Table \ref{jetcase}) of collimated fireball quantities in terms of rotation 
parameter $a^2\Omega^2_c\propto J/M$ at the end of homologous collapses. 
From these numerical trends, we infer collimated fireball intrinsic correlations, 
including the collimated fireball angle $\theta_{\rm coll}$. 
We omit the superscript $^j$ of collimated fireball quantities for simplifying notations unless otherwise specified.

\subsection{Intrinsic correlations imprint on observed GRB data}\label{pcon}

The theoretically obtained intrinsic correlations (\ref{chi}) and (\ref{chil}) seem to give the Amati and Yonetoku 
empirical correlations in GRBs \cite{Amati2002,Yonetoku2004}, if we identify the photon-pair collimated fireball energy $E_\gamma$ and mean energy $\langle T_\gamma \rangle$ to the observed isotropic equivalent energy $E_{\rm iso}$ and peak energy $E_p$. 
However, we cannot simply 
make such an identification for the reasons that a 
photon-pair collimated fireball is opaque and undergoes complex processes up to the transparency at which it becomes an observably relevant photosphere or collimated one. One of the reasons is that the peak energy $E_p$ value sensitively depends on the baryonic material loading (ejecta mass)
$M_{\rm ej}$ on collimated fireballs, 
see for example, Refs.~\cite{Ruffini1999,Ruffini2000,Meszaros2000,Zhang2021}.  
However, we present a general discussion on the connections between theoretical correlations (\ref{chi}-\ref{time3}) and trends (Table \ref{jetcase}) of different collimated fireballs and observational correlations of different GRB sources.

The collimated fireballs provide the energetic characteristics for observed GRB events. Suppose each collimated fireball created by a gravitational collapse corresponds to a GRB event/source observed. An initially opaque collimated fireball ({\it initial state}) undergoes complex ({\it intermediate processes}) to a transparent ``photosphere'' ({\it final state}) of GRB phenomena observed. The collimated fireball quantities $T_\gamma$, $E_\gamma$, $L_\gamma$ and $R_\gamma$ are not the same as the observed GRB event's mean energy, total energy, total luminosity, and time duration. Therefore, the theoretical correlations (\ref{chi}-\ref{time3})  and trends (Table \ref{jetcase}) cannot be examined by directly comparing them with the observed quantities in GRB events.

However, collimated fireball overall quantities ($T_\gamma, E_\gamma, L_\gamma,
R_\gamma$), intrinsic relations (\ref{chi}-\ref{time3}) and trends (Table \ref{jetcase}) must {\it imprint (code)} themselves in observed GRB data. Although the intermediate processes from a collimated fireball to an observed GRB source are complex, they {\it randomly} differ from one GRB source to another. Such randomness implies that the observed GRB 
quantities, corresponding to collimated fireball counterparts, should randomly scatter around the intrinsic relations (\ref{chi}-\ref{time3}) and trends (Table \ref{jetcase}) if these scaling relations and trends truly reflect the universal natures of centre engines for GRB energetics. 
Let's take into account more data on GRB events. We will achieve a more statistically confident level on the validity of the intrinsic relations (\ref{chi}-\ref{time3}) and trends (Table \ref{jetcase}) of collimated fireball quantities. 
These are only general discussions on the observational 
relevance of the intrinsic relations (\ref{chi}-\ref{time3}) and trends (Table \ref{jetcase}) in the statistical sense 
by using a large sample 
of observed GRB data. For this purpose, the relevant GRB phase identification, spectrum and event selection, and elaborate data are required.
These subjects are currently under investigation.

We end this section by emphasizing that the GRB physics is most complex as various dynamic processes are involved in time evolution. The central engine physics at the initial explosion could be simpler, but masked by surrounding opaque material, cannot be directly probed by observations. In theoretical correlations (\ref{chi}-\ref{time3}), the universal scaling laws should be primary, and the relations between scaling indexes can receive corrections. Therefore, the verifications of theoretical correlations (\ref{chi}-\ref{time3}) and correlations relating to jet angles discussed below must be conducted carefully with caution. They cannot be justified or falsified by a limited number of observations. We like to mention recent empirical studies \cite{Dainotti2020, Dainotti2022} of GRBs' luminosity and duration correlation, similar to Eq.~(\ref{time3}) or Eq.~(7.7) of Ref.~\cite{Xue_2021}.

\begin{table*}
\centering
\begin{tabular}{cccccccc}
$a^2\Omega^2_c$ &$\theta_{\rm coll}$ & $E^j_p/E_p$ &$E^j_{\gamma}/E_{\gamma}$&$L^j_{\gamma}/L_{\gamma}$ & $R^j_{\gamma}/R_{\gamma}$ & $N^j_{\gamma}/N_{\gamma}$ &$V^j_{\gamma}/V_{\gamma}$\cr
\hline
\hline
$0.001$&$13.2^\circ$ & $1.01$& $0.998$ & $4.29$ & $6.9/30$ & $0.99$& $0.92$\cr
$0.005$ &$7.8^\circ$ & $1.25$& $0.99$ & $7.24$ & $4.1/30$ & $0.97$& $0.63$\cr
$0.01$&$6.3^\circ$ & $1.38$& $0.98$ & $8.90$ & $3.3/30$ & $0.96$& $0.52$\cr
$\Rightarrow 0.02$&$4.8^\circ$ & $1.52$& $0.97$ & $11.6$ & $2.5/30$ & $0.93$& $0.42$\cr
$0.03$ &$4.2^\circ$& $1.60$& $0.95$ & $12.9$ & $2.2/30$& $0.91$& $0.37$\cr
$0.04$ &$3.8^\circ$ & $1.66$& $0.94$ & $14.1$ & $2.0/30$& $0.89$& $0.33$\cr
$0.05$ &$3.4^\circ$ & $1.71$& $0.93$ & $15.5$ & $1.8/30$& $0.88$& $0.31$\cr
$0.06$ &$3.2^\circ$ & $1.74$& $0.92$ & $16.2$ & $1.7/30$& $0.86$& $0.29$\cr
$0.07$ &$3.1^\circ$ & $1.78$& $0.91$ & $17.1$ & $1.6/30$& $0.85$& $0.27$\cr
$0.08$ &$2.9^\circ$ & $1.81$& $0.90$ & $18.0$ & $1.5/30$& $0.84$& $0.26$\cr
$0.09$ &$2.7^\circ$& $1.84$& $0.89$ & $19.1$ & $1.4/30$& $0.83$& $0.25$\cr
$0.10$&$2.5^\circ$& $1.86$& $0.88$ & $20.3$ & $1.3/30$ & $0.82$& $0.24$\cr
$0.15$ &$2.1^\circ$ & $1.95$& $0.85$ & $23.2$ & $1.1/30$ & $0.78$& $0.20$\cr
\hline
\hline
\end{tabular}
\caption{For selected values of the rotation parameter $a^2\Omega^2_c$, which is proportionally related to the ratio $J/M$ (\ref{jmr}), we tabulate the collimated angle $\theta_{\rm coll}$ (\ref{jeta}) and the ratios of the spectral peak $E^j_p/E_p$, total energy $E^j_{\gamma}/E_{\gamma}$, maximal luminosity $L^j_{\gamma}/L_{\gamma}\propto (E^j_\gamma/E_\gamma)(R_\gamma/R^j_\gamma)$, minimal size $R^j_{\gamma}/R_{\gamma}$, total particle number $N^j_{\gamma}/N_{\gamma}$ 
and volume
$V^j_{\gamma}/V_{\gamma}$ between the photon-pair spherical and collimated fireballs for comparisons. The collimated (with superscript $j$) and spherical fireball quantities are given by Eqs.~(\ref{Rsize},\ref{totn},\ref{atem}). Their ratios do not explicitly 
depend on the initial (final) core centre density $\rho_c^{\rm in}$ ($\rho_c^{\rm fi}$) of gravitational collapses.}\label{jetcase}
\end{table*}

\section{Long vs short GRBs by progenitor angular momentum}
\label{spherevsjet}

Based on the numerical results of 
Eqs.~(\ref{Yn}-\ref{atem}) 
and Fig.~\ref{sphere-jet1} for the parameters $\it Blue$ case in the list (\ref{para}), using the initial condition (\ref{inrho}), 
we calculate the photon-pair collimated fireball angle $\theta_{\rm coll}$, spectral peak $E^j_p$, total energy $E^j_\gamma$ and luminosity $L^j_\gamma$ in comparison with the sphere counterparts. We select the rotation parameter $a^2\Omega^2\propto J/M$ values to gain an insight into the trends of how collimation geometric and energetic features vary with the ratio of total mass and angular momentum $J/M$ in Table \ref{jetcase}.

\subsection{Intrinsic trends of collimated fireball properties varying with $J/M$}\label{trends}

Table \ref{jetcase} from left to right columns shows the trends of collimated fireball quantities 
varying as the angular momentum and mass ratio $J/M\propto a^2\Omega^2_c$ increases, 
\begin{enumerate}[(1)]
\item the collimated fireball angle $\theta_{\rm coll}$ becomes narrower; 
\item the fireball energy spectrum $E^j_p$ becomes harder; 
\item the fireball isotropic equivalent energy $E^j_\gamma$ becomes smaller;
\item the fireball isotropic equivalent luminosity $L^j_\gamma$ becomes larger; 
\item the fireball characteristic size $R^j_\gamma$ 
(time $\tau^j_\gamma=R^j_\gamma/c$) becomes smaller (shorter).
\end{enumerate}
We obtain these numerical results
for massive slowly-rotating core collapses with the ratios $J/M\ll 1$. 

Nevertheless,
we extrapolate these trends in Table \ref{jetcase} to the 
larger $J/M$ values for fast-rotating cores and binary mergers, 
to gain an insight into the collimated fireballs generated in gravitational collapses of fast-rotating systems. The arguments are 
as follows:
\begin{enumerate}[(i)]
\item The microscopic physics processes of hadron collisions (Sec.~\ref{hadron}) accounting for 
producing fireballs in slow- or 
fast-rotating massive collapsing cores and binary coalescence are the same. It occurs at nucleon interaction time and length scales when rotating massive cores and binary mergers run into the homologous collapsing phase around the nuclear density. 
The collimated fireball quantities are mainly determined by the initial centre density $\rho_c^{\rm in}$ 
of massive collapsing cores or the effective values $(\rho_c^{\rm in})_{\rm eff}$ 
of binary mergers. These microscopic processes at length scales of nuclear interactions play essential roles in fireballs' formation.  
\item 
The macroscopic physics processes of the dynamic interplay
between gravitational binding and rotational repelling energy for collimating fireballs is the same for slow- or fast-rotating massive core collapses and binary mergers. 
The processes and results are characterized by the 
ratio $J/M$ of total angular momentum 
$J$ and mass $M$ of systems, which are conserved in the 
entire core collapsing or binary merging process. Such an overall ratio 
$J/M$ possesses the same physical meaning for a rotating core collapsing and binary merging into a rotating core collapsing near the nuclear density when fireballs occur. 
Therefore, the $J/M$-described macroscopic processes should play a similar key role in fireballs' collimation for massive core collapses and binary mergers. 
\item In addition, the obtained numerical trends of collimated fireball characteristic quantities vary {\it monotonically} in terms of $J/M$, following the expected physical behaviours from the dynamic
interplay between gravitational binding and rotational repelling
energy for collimating fireballs.
\item Numerical simulations show
binary coalescence has a very different structure from core collapses. In binary coalescence cases, collimated
fireballs should form in the latest 
gravitational collapsing phase of binary coalescence when binary stellar cores merge, becoming a 
fast-rotating core collapsing around the nuclear density. From these insights, we expect some similarity 
of fireballs' collimation between core collapses and binary mergers in terms of their $J/M$. To demonstrate it, full numerical simulations in the GR framework are necessary.
\end{enumerate}

However, these arguments are highly speculative rather than justified.
We cannot conclude that the trends apply to fast-rotating collapsing systems and binary mergers relating to GRBs' observations because of our analytical studies based on a 
simplified model of a slowly rotating and massive core collapsing in the Newtonian approximation. The key issue is how fireballs collimate in gravitational collapses of 
fast-rotating systems. It is very complicated to perform analytical studies on collimated fireballs formed in the gravitational collapses of fast-rotating massive cores and binary systems in the GR framework. To
obtain conclusively the properties of collimated fireballs varying in terms of $J/M$, one has to perform numerical studies in the GR framework to study fast-rotating collapsing systems and binary mergers.

Despite our lack of such numerical simulations, 
it is worthwhile to examine the 
self-consistency of these theoretically speculated $J/M$ trends (Table \ref{jetcase}) of collimated fireballs from the empirical viewpoint of long and short GRBs' observations.

\subsection{Long vs short GRBs and jet angles}

These collimated fireball quantities 
and trends in Table \ref{jetcase} cannot directly connect with observed quantities in GRBs data. However, as discussed in Sec.~\ref{pcon}, these intrinsic trends 
(1)-(5) 
of energetic and geometric quantities varying from nearly spherical to collimated fireballs in terms of $J/M\propto a^2\Omega_c^2$ 
must {\it imprint or code} in the data of observed GRBs if these trends truly reflect the universal nature of collimated fireball energetics of GRB centre engines. In other words, the imprints of these trends on data should be significantly evident in a large sample of observed GRB data.

To see if the trends (1)-(5) and Table \ref{jetcase} are relevant to observations connecting with jet features,
we need to assume the collimated fireball angle $\theta_{\rm coll}$ proportional relation (or positive correlation) (\ref{jeta1}) to the jet angle $\theta_{\rm jet}$ and observational one $\theta^*_{\rm jet}$, as discussed in Sec.~\ref{cangle}.
Based on the previous discussions on 
small $J/M$ for core collapses 
and large $J/M$ for binary mergers, 
we will examine the trends (1)-(5) in Table \ref{jetcase} as $J/M$ 
increase to see whether or not the trends and classifications by $J/M$ are consistent with observational data 
analysis of short vs long GRBs by a comprehensive study of the energetics and prompt gamma-ray correlations \cite{2015MNRAS.451..126S}.

We observe that the trends (1)-(5) of collimated fireball quantities as increasing angular momentum and mass ratio $J/M$ in Table \ref{jetcase} support the common view that long GRBs from massive rotating core collapses (small $J/M$) and short GRBs from massive binary coalescence (large $J/M$).  Based on the positive correlation $\theta_{\rm coll}\sim \theta_{\rm jet}\sim \theta^*_{\rm jet}$ (\ref{jeta1}), 
we find some novel features on observed ultra-relativistic jet angles $\theta^*_{\rm jet}$ correlating to long and short GRBs.
The following qualitative observations and discussions are in order.
\begin{enumerate}[(i)]
\item Due to the angular momentum $J/M$ and repulsive centrifugal potential increase, the gravitational binding energy gain becomes smaller. The ratios $E_\gamma^j/E_\gamma$ and $N_\gamma^j/N_\gamma$ are smaller than one but still
in the order of unity, because most photons and pairs are produced at the centre ($z=0, y=0$) of the collimated fireball, see Figs.~\ref{sphere-jet} and \ref{sphere-jet1}.

\item The collimated fireball spectral peak $E_p^j/E_p$ significantly increases up to twice as large as the spherical one, because the photon-pair energy density increases as the collimated fireball volume $V^j$ decreases. As a result, the collimated fireball spectral peak $E^j_p$ increases, the total energy $E^j_\gamma$ and the characteristic time scale 
$R^j/c$ decrease as angular momentum $J/M$ increases. 
These trends might give a possible explanation for two distinct Amati correlations, relating to $E_p\propto E_\gamma^\chi$ (\ref{chi}),
respectively for long and short bursts' data scattering on the $E_p$-$E_\gamma$ plane, see Fig.~7 (right) of the reference \cite{2015MNRAS.451..126S}, where two distinct Amati relations are observed for long and short bursts.

\item 
The collimated fireball spectral peak $E^j_p$ and the peak luminosity $L^j_\gamma$ increase as the angular momentum $J/M$ increases. It implies the Yonetoku correlation, relating to $E_p\propto L_\gamma^{\chi/(\chi+1)}$ (\ref{chile}), is not the same for long and short bursts' data scattering on the $E_p$-$L_\gamma$ plane, as shown in Figure 7 (left) of the reference \cite{2015MNRAS.451..126S}.

\item The collimated and sphere fireball sizes' ratio  $R_\gamma^j/R_\gamma \sim {\mathcal O}(10^{-1})$. It implies that long bursts duration $T_{90}\sim R_\gamma/(2c\Gamma^2_0)$ (\ref{t90}) are about 10 times lager than short bursts $T^j_{90}\sim R^j_\gamma/(2c\Gamma_{0j}^2)$. 
Here, we assume hydrodynamic expansion ejecta Lorentz factors $\Gamma_0/\Gamma_{0j}\sim {\mathcal O}(1)$, considering the obtained result $E^j_\gamma/E_\gamma\sim {\mathcal O}(1)$ in Table \ref{jetcase} and assuming ejecta mass $M_{\rm ej}$ is the same order of magnitude in spherical and collimated cases, i.e., $M^j_{\rm ej}/M_{\rm ej}\sim {\mathcal O}(1)$. The ratio $T^j_{90}/T_{90} \sim {\mathcal O}(10^{-1})$ is qualitatively consistent with two distinct classes of long (soft) and short (hard) bursts' time-duration data scattering on the $E_p$-$T_{90}$ plane, see Figures 1 and 6 (up-right) of the reference \cite{2015MNRAS.451..126S}. 

\item 
The trends of numerical results as increasing $J/M$ in Table \ref{jetcase} imply possibly positive or negative correlations between the collimated fireball angle $\theta_{\rm coll}$, burst duration $T_{90}^j$, spectral peak $E_p^j$, total energy $E_\gamma^j$, luminosity $L_\gamma^j$.
These trends indicate that short GRBs (hard photon spectrum) tend to be emitted from a narrow collimated fireball, in comparison with long GRBs (soft photon spectrum) preferentially come from a nearly 
spherical fireball. In other words, short GRBs from binary coalescence of large angular momenta $J/M$ are more jetting (smaller $\theta_{\rm coll}\sim \theta_{\rm jet}$) than long GRBs from rotating core collapses of small angular momenta. Short GRBs originate from more collimated fireballs with large angular momentum 
$J_\gamma/M$. Long GRBs originate from less collimated fireballs with small angular momentum 
$J_\gamma/M$.

\item The above observations on the trends vs angular momentum $J/M$ and Sec.~\ref{sepx} discussions on the separatrix $(J/M)_{\rm sepa}$ indicate that short GRB progenitors possess larger $J/M$ values than long GRB progenitors. The separatrix (\ref{separa}) of this oversimplified model cannot infer the separatrix $T_{90}\approx 1\sim 2$ seconds between long GRBs (massive core collapses) and short GRBs (binary coalescence). However, following the discussions on the viewpoint (iv)
above, we find that $R^j_\gamma/R_\gamma=\tau^j_\gamma/\tau_\gamma \approx 10^{-1}$ corresponds to the line $a^2\Omega_c^2 \approx 0.02$ indicated by ``$\Rightarrow$''
in Table \ref{jetcase}. It infers collimated fireball angles $\theta_{\rm coll} < 5^\circ$, thus observed jet angles $\theta^*_{\rm jet} < 5^\circ$  
for the cases of short GRBs originating from massive binary coalescence. Analogously to the $T_{90}$ separatrix, the separatrix of GRB jet angles should be in short and long GRB data. This also implies that short and long GRB events should have a bimodal distribution in terms of jet angles \footnote{Thanks to the Referee for this comment.}.

\item 
The above observations are based only on the features of collimated fireballs in the initial prompt era. The GRBs' ultra-relativistic outflow
jet angles $\theta_{\rm jet}\approx \Gamma^{-1}_{\rm jet}\propto (E_\gamma/M_{\rm ism})^{-1/8}$ determined by the observed jet breaks in the light curves of the afterglow era \cite{Sari1999, Frail:2001qp} for spherical adiabatic evolution, when jets propagate and interact with the surrounding interstellar medium (ISM). The $M_{\rm ism}$ is the total ISM mass engulfed by propagating jets up to the jet breaks' time $t_{\rm jet}\propto M_{\rm ism}^{1/3}$, and energy-momentum conservation laws are used to 
determine how the Lorentz factor $\Gamma_{\rm jet}$ decreases from $\Gamma_0$ (\ref{Gamma0}). We assume that jets conserve the angular momentum along their direction by propagating through a homogeneous ISM without shearing forces. Short GRBs jets from binary coalescence have large $\Gamma_0$, small $\theta_{\rm jet}$ and cross-section interacting with ISM, while long GRBs jets from massive core collapses have small $\Gamma_0$, large $\theta_{\rm jet}$ and cross-section interacting with ISM. We need to collect more data on jet breaks, particularly for short GRBs events, to see if there is any bimodal distribution in terms of jet angles $\theta_{\rm jet}$, 
and whether it is consistent with 
the $T_{90}$ and $E_p$ dichotomies for short and long GRBs.   

\end{enumerate}

We have not found inconsistencies between the theoretical scenario and observational phenomena, although the adopted model and approximation are oversimplified for complex GRB progenitor systems. However, 
the model adopted is still far from a realistic application in analyzing GRB progenitor properties and their consequences. Despite these weaknesses, these scenario insights underscore the importance of considering the angular momentum of rotating systems in collapses and collimated fireball dynamics in studying GRBs' spectrum, energy, luminosity, duration, jetting angle, and their correlations.

\subsection{Angular momentum separatrix of 
core collapses and binary mergers}\label{sepx}

In addition, the distinct separation between long and short GRB classes is evident in observations. We attempt to give a possible explanation of this phenomenon consistently with our arguments on the properties of GRBs' jet progenitors varying with their $J/M$ in Sec.~\ref{trends}.
For larger $J/M$ values and rotational repelling energies $U_c/|U|$ (\ref{ejgur}), a massive rotating core cannot maintain the equilibrium conditions (\ref{euler1T}) with given material binding energy $\lambda$. Thus, rotating stellar cores split.    
In this case, the most probable configuration is a binary system. This condition indicates two distinct classes: massive rotating core collapses with small $J/M$ values and binary mergers with large $J/M$ values.

We speculate that 
the separatrix between two distinct classes occurs when the equilibrium condition (\ref{euler1Teq}) or (\ref{euler1T}) breaks down. Namely, the centrifugal potential $\phi_c$ and heat function $h$ are too large to be balanced by the gravitational binding potential $\phi$ and the material binding energy $\lambda$. Note that the $\lambda$ value should not depend on whether the system is a single core or a binary object at the nuclear density. 

Considering the particular case that 
the heat function $h$ balances the gravitational binding
potential $\phi$ in the equilibrium condition (\ref{euler1Teq}), we find the critical centrifugal potential balanced by material binding potential energy $\lambda$, 
\begin{eqnarray}
(\phi_c)_{\rm sepa}\approx \frac{(v_s^c)^2}{\gamma-1}\frac{\lambda}{6} r^2_s\quad \Rightarrow \quad  (a^2\Omega^2)_{\rm sepa}\sim \frac{\lambda}{3(\gamma-1)},
\label{separa}
\end{eqnarray}
where $\phi_c\approx \frac{(v_s^c)^2}{2} r^2_sa^2\Omega_c^2$ (\ref{centrifugal}) by setting 
$r=r_s$ and $\sin^2\theta =1$. 
The critical condition 
(\ref{separa}) indicates a separatrix $(J/M)_{\rm sepa}$ via the angular-momentum relation (\ref{jmr}) or a separatrix $(U_c/|U|)_{\rm sepa}$ via the rotation-energy relation (\ref{ejgur}).
This separatrix implies the following two separate situations: 
\begin{enumerate}[(a)]
\item
When the material binding energy is larger than the repulsive rotation energy, all materials bind together and undergo massive core collapses. 
This case is for slowly rotating cores $J/M\ll 1$ or $U_c/|U|\ll 1$. Such progenitors produce less collimated fireballs with small angular momentum $J_\gamma/M\ll 1$.
\item 
On the contrary, all materials cannot bind together as a single massive core in collapses when the systems have large $J/M$
or $U_c/|U|$ values, and the material binding energy is smaller than the repulsive rotation energy.
Therefore, some gravitational binding energy is required 
to balance rotational energy. Such systems are very probably realized as binary systems undergoing gravitational mergers. Such progenitors produce more collimated fireballs with large angular momenta $J_\gamma/M$.
\end{enumerate}

Although we are not able to calculate the separatrix, these discussions give an insight into the distinction between long and short GRB classes relating to the angular momentum and mass ratios $J/M$ of massive rotating core collapses and binary 
coalescence. 
The realistic situation of separating two distinct GRB classes should be much 
more complex. Our theoretical model and analysis methods are too simplified and approximate for fast-rotating core collapses and binary coalescence. 
Therefore, our physical explanations of jet angle correlation, short GRBs and long GRBs distinction in terms of $J/M$ are qualitative and speculative. Further theoretical and empirical studies on these explanations are necessary and welcome.

\section{Summary and remarks on GRB light angular momenta}\label{end}

In summary, to understand the physics of progenitors generating highly collimated GRB events, we adopt  
a simplified model in the Newtonian approximation to qualitatively study photon-pair collimated fireballs generated by the gravitational collapse of slowly rotating 
massive cores and binary systems. 
Our qualitative studies infer the following photon-pair collimated fireball features: (i) collimated fireballs acquire nontrivial angular momenta $J_\gamma$ from massive rotating 
systems (progenitors) in gravitational collapses and make ultra-relativistic jets due to the angular momentum $J_\gamma$ conservation along propagating direction; (ii) the progenitor angular momentum and mass ratio $J/M$ determines the collimated fireball angle $\theta_{\rm coll}$ relating to observed GRB jet angles; 
(iii) collimated fireball quantities ($\theta_{\rm coll}$, peak spectra, total energy, luminosity, and time scale) have positive and negative correlations among them, whose imprints on observational quantities can be examined by using a large sample of observed GRB data; (iv) the distinctions between short and long GRBs' properties are due to the different ratios of the angular momenta and masses of systems that undergo gravitational processes of massive binary coalescence or core collapse. These highlight the significant role of progenitor angular momentum and mass ratios in determining the collimated fireball characteristics and, consequently, the observable properties of GRBs, paving the way for a more comprehensive understanding of GRB jet progenitors. 

On the one hand, more theoretical studies and numerical simulations of massive rotating core collapse and binary coalescence 
are necessary to understand better how GRB progenitors generate collimated fireballs with nontrivial angular momenta along the rotating direction of collapse systems, 
and such collimated fireball outflows launch ultra-relativistic jets, propagating with angular momentum conservation along jet directions. 
These studies from the first principle will help refine our understanding of the dynamics of collimated fireball formation, ultra-relativistic jet launching and propagation, and resultant GRB properties. 

On the other hand, it is important to perform the data analyses on empirical correlations of short and long GRB data, connecting to collimated fireball characteristics: energetic, geometric, and temporal features. These data analyses will reveal the intrinsic correlations among collimated fireball characteristics, thus providing deeper insights into the jet progenitor systems and the physical mechanisms driving GRBs. 

To end this article, we mention that emitted from an ultra-relativistic jet with nontrivial angular momentum $J_\gamma$ along the axis of propagation and propagating in external magnetic fields, GRB light should possibly carry nontrivial spin angular momentum (SAM) and orbital 
angular momentum (OAM). The former leads to circular wave polarization, and the latter makes the helical twisting wavefront. These issues could be interesting for GRB theoretical studies and
observational analyses.



\providecommand{\href}[2]{#2}\begingroup\raggedright\endgroup

\end{document}